\documentclass[]{aa}

\usepackage{txfonts}
\usepackage{graphicx}
\usepackage{amsmath}

\begin{document}

\title{In and out star formation in $z\sim1.5$ quiescent galaxies from rest-frame UV spectroscopy and the far-infrared}

\author{R. Gobat\inst{1}
\and E. Daddi\inst{2}
\and V. Strazzullo\inst{3}
\and B. Garilli\inst{4}
\and M. Mignoli\inst{5}
\and Z. Ma\inst{6}
\and S. Jin\inst{2,7}
\and C. Maraston\inst{8}
\and G. Magdis\inst{9,10}
\and M. B\'{e}thermin\inst{11}
\and M. Cappellari\inst{12}
\and M. Carollo\inst{13}
\and A. Cimatti\inst{14}
\and C. Feruglio\inst{15}
\and M. Moresco\inst{14}
\and M. Onodera\inst{16,17}
\and L. Pozzetti\inst{5}
\and A. Renzini\inst{18,16}
\and M. Sargent\inst{19}
\and F. Valentino\inst{2}
\and A. Zanella\inst{2}
}

\institute{
School of Physics, Korea Institute for Advanced Study, Hoegiro 85, Dongdaemun-gu, 
Seoul 02455, Republic of Korea
\and Laboratoire AIM-Paris-Saclay, CEA/DSM-CNRS--Universit\'{e} Paris Diderot, Irfu/Service 
d'Astrophysique, CEA Saclay, Orme des Merisiers, F-91191 Gif sur Yvette, France
\and Department of Physics, Ludwig-Maximilians-Universit\"{a}t, Scheinerstr. 1, D-81679 
M\"{u}nchen, Germany
\and INAF-IASF, Via Bassini 15, I-20133 Milano, Italy
\and INAF-Osservatorio Astronomico di Bologna, Via Ranzani 1, I-40127 Bologna, Italy
\and Department of Astronomy, University of Science and Technology of China, Hefei 230026, China
\and Key Laboratory of Modern Astronomy and Astrophysics in Ministry of Education, School of Astronomy 
and Space Science, Nanjing University, Nanjing 210093, China 
\and Institute of Cosmology and Gravitation, Dennis Sciama Building, Burnaby Road, Portsmouth PO1 
3FX, UK
\and Dark Cosmology Centre, Niels Bohr Institute, University of Copenhagen, Juliane Mariesvej 30, 
DK-2100 Copenhagen, Denmark
\and Institute for Astronomy, Astrophysics, Space Applications and Remote Sensing, National 
Observatory of Athens, GR-15236 Athens, Greece
\and European Southern Observatory, Karl-Schwarzschild-Strasse 2, D-85748 Garching, Germany
\and Department of Physics, University of Oxford, Keble Road, Oxford OX1 3RH, UK
\and Institute for Astronomy, ETH Zurich, CH-8093 Zurich, Switzerland
\and Department of Physics and Astronomy, University of Bologna, Viale Berti Pichat 6/2, I-30127 
Bologna, Italy
\and INAF-Osservatorio Astronomico di Trieste, Via G.B. Tiepolo 11, I-34143 Trieste, Italy
\and Subaru Telescope, National Astronomical Observatory of Japan, National Institutes of Natural 
Sciences (NINS), 650 North A'ohoku Place, Hilo, HI 96720, USA
\and Department of Astronomical Science, SOKENDAI (The Graduate University for Advanced Studies), 
650 North A'ohoku Place, Hilo, HI 96720, USA
\and INAF-Osservatorio Astronomico di Padova, Vicolo dell'Osservatorio 5, I-35122 Padova, Italy
\and Astronomy Centre, Department of Physics and Astronomy, University of Sussex, Brighton, BN1 9QH, 
UK
}

\date{}

\abstract{
We present a sample of 34 spectroscopically confirmed \emph{BzK}-selected $\sim10^{11}$~M$_{\odot}$ 
quiescent galaxies (pBzK) in the COSMOS field. The targets were initially observed with VIMOS on 
the VLT to facilitate the calibration of the photometric redshifts of massive galaxies 
at $z\gtrsim1.5$. Here we describe the reduction and analysis of the data, and the 
spectrophotometric properties of these pBzK galaxies. In particular, using a spatially resolved 
median 2D spectrum, we find that the fraction of stellar populations with ages $<$1~Gyr is at 
least 3 times higher in the outer regions of the pBzK galaxies than in their cores. This results in 
a mild age gradient of $\Delta \text{age}\leq0.4$~Gyr over $\sim$6~kpc and suggests either the 
occurrence of widespread rejuvenation episodes or that inside-out quenching played a role in the 
passivization of this galaxy population. 
We also report on low-level star formation rates derived from the [OII]3727$\AA$~emission line, 
with $\text{SFR}_{\text{OII}}\sim3.7-4.5$~M$_{\odot}$~yr$^{-1}$. This estimate is confirmed 
by an independent measurement on a separate sample of similarly-selected quiescent galaxies in the 
COSMOS field, using stacked far-infrared data 
($\text{SFR}_{\text{FIR}}\sim2-4$~M$_{\odot}$~yr$^{-1}$). This second, photometric sample also displays 
significant excess at 1.4~GHz, suggestive of the presence of radio-mode AGN activity.
}


\titlerunning{VIMOS spectroscopy of pBzK galaxies}
\authorrunning{Gobat et al.}

\maketitle

\section{Introduction}

Since the passing of the cosmic high noon and the peak of star formation in the Universe, the most 
massive galaxies have been largely quiescent, with low to undetectable rates of star formation 
and so-called early-type morphologies. Their evolution since $z\sim1$~has been traced and 
extensively investigated for well over a decade \citep[e.g.,][]{Kau04,Tho05,Gal06}.  At higher 
redshift, and especially above $z\sim1.5$, quiescent galaxies form a conspicuous minority population 
\citep[e.g.,][]{Cim04,MCy04,Dad05}, now often used as a tracer of overdense structure 
\citep{Gob11,Spi12,Chi14,Str15}. 
They have so far been spectroscopically confirmed up to $z\sim3$~\citep{vDk08,Gob12,Hil16}, with 
adequate ($\gtrsim10$) samples now existing up to 
$z\sim2$~\citep[e.g.,][]{Tan13,Whi13,Gob13,Bel14,New14,Kro14}. Their appearance only $\sim$2~Gyr 
after the Big Bang, as well as their ages and abundances at $z>1.5$, imply that their progenitors 
must have converted their gas into stars on short timescales and that the suppression of star 
formation happened relatively rapidly \citep{Ono15}. 
Furthermore, observational data at both low and high redshift suggest that the timescale for the 
quenching of star formation in massive ($\gtrsim10^{11}$~M$_{\odot}$)~galaxies does not depend on 
the local environment but is mostly set by processes internal to their host halo 
\citep[e.g.,][]{Bal06,Tho10,Pen10,Gob13,New14}. This so-called ``mass quenching'' can in principle 
be mediated by a variety of unrelated mechanisms, such as: virial shock heating of infalling gas in 
$>10^{12}$~M$_{\odot}$ halos \citep[commonly referred to as halo quenching]{Bir03,Dek06}, 
stabilization of gas through bulge growth 
\citep[or inside-out, sometimes also called morphological, quenching;][]{Mrt09}, and 
various feedback processes from either the central nucleus \citep[e.g.,][]{Gra04,DiM05,Crt06,Hop06} 
or star formation \citep[e.g.,][]{Cev09,Fel15} which can directly expel the gas from the galaxies 
or contribute to the heating of halo gas. However, whether one of these candidate mechanisms is 
truly dominant during the initial quenching, or whether the suppression of star formation in 
massive galaxies requires a combination of factors \citep[e.g.,][]{Woo15}, is still not well 
understood.\\

On the other hand, these various quenching processes differ in their timescales and some mechanisms 
(radio-mode AGN feedback, virial shock heating) are expected to leave in quenched galaxies a 
substantial amount of gas \citep[e.g.,][]{Gab10} that can sustain some star formation activity 
long after the initial quenching event. While the direct detection of gas reservoirs in high-redshift 
quiescent galaxies remains on the edge of instrumental feasibility \citep{Sar15}, constraints on recent 
and ongoing star formation can be obtained from other indicators. 
For example, clues on the star formation history (SFH) prior to quenching can be derived from 
optical spectroscopy, although such analysis requires a good signal-to-noise ratio (S/N) and is 
therefore usually performed on $z<1$~galaxies \citep[e.g.,][]{Cho14,Gal14,She15,Cit16}. At higher redshifts, 
composite near-infrared (NIR) spectra are typically needed \citep[e.g.,][]{Ono12,Ono15,Whi13,Lon14}. When 
these are not available, rest-frame ultraviolet (UV) light can function as a loose proxy of recent star 
formation \citep[e.g.,][]{Ret10,Ret11}, while more recently deep mid-infrared (MIR) and far-infrared 
(FIR) observations have been used to estimate the residual star formation rate (SFR) of the massive 
$z>1$~quiescent population \citep{Wag15,Man16}. On the other hand, this latter approach 
is limited by the fact that neither the UV or IR indicators used are uniquely correlated with 
ongoing star formation, but can also originate from old stellar populations \citep[e.g.,][]{Fum14}, 
which makes the results of a purely photometric analysis ambiguous.\\

In this paper, we present an analysis of spectroscopic observations of a sample of massive, 
\emph{BzK}-selected quiescent galaxies in the COSMOS field \citep{Sco07}. These were originally obtained 
to help optimize photometric redshifts for $z\gtrsim1.5$~passive galaxies, and a comparison of spectroscopic 
and photometric redshifts for this sample has already been presented in \citet{Str15}. Furthermore, an analysis 
of their spectral energy distributions (SED) is described in \citet{Cpo16}. Here we report on stellar 
populations properties and star formation rates from both spectroscopic and FIR tracers.
This paper is structured as follows: In Section \ref{sample} we present the sample, the observations, 
and describe the method used to derive spectroscopic redshifts; we detail our spectroscopic analysis 
in Sections \ref{stelpop} and \ref{sfr}; we discuss its implications in Section \ref{disc}, and 
summarize our conclusions in Section \ref{sum}. 
A table with the spectroscopic redshifts of this sample is given in Appendix \ref{ztable}. We 
assume a $\Lambda$CDM cosmology with $H_0=70$~km~s$^{-1}$~Mpc$^{-1}$, $\Omega_{\text{M}}=0.27$, 
and $\Omega_{\Lambda}=0.73$, and all cited quantities assume a \cite{Sal55} initial mass function 
(IMF).

\section{\label{sample}Sample selection and data reduction}

We selected galaxies from the catalog of \citet{McC10}, identified as passive using the \emph{BzK}~criterion 
\citep{Dad04}, and brighter than $I_{\text{AB}}=25$, so as to be detectable in at most a few nights of 
spectroscopic observations at optical/NIR wavelengths.
This criterion yielded $\sim$2700 passive BzK (pBzK) galaxies over the full COSMOS field, from which we 
selected an area containing 78 pBzKs. Among these galaxies, we targeted 29 pBzKs that were undetected at 
24~$\mu$m in the 2010 catalog. The configuration of the spectroscopic mask allowing it, we included 
another 6 pBzKs that were detected in \emph{Spitzer}/MIPS imaging, for a total of 35 targets. 
This corresponds to 56\% of pBzKs with $\log$$M_{\star}>10.8$~in the target area, and 90\% of those with 
$\log$$M_{\star}>11.3$. 
These galaxies were observed with VIMOS on the VLT in service mode, during ESO programs 086.A-0681 and 
088.A-0671 (February-March 2011 and February-March 2012, respectively; PI E. Daddi). The total exposure 
time per field was 10.5 hours, split into 72 exposures of 523 seconds each.  
We used the medium resolution grism, with a dispersion of 2.5~$\AA$~per pixel (0.205'' in the spatial 
direction), coupled to the GG475 order-sorting filter which limits the wavelength range to 
$4500-10000$~$\AA$, and a slit width of 1'' corresponding to a resolution of $\text{R}=580$. 
To allow for better background subtraction, we applied the `nod along slit' technique, with a shift of 
2'' between one exposure and another, for a total of 5 positions. The airmass during the observation was 
$\sim$1.15, while the typical seeing was better than 1''.
The data were then reduced using the VIPGI pipeline \citep{Sco05} as follows: the position of each 2D 
spectrum was first determined for each single exposure using the flat field; sky subtraction was then 
performed on the bias-subtracted raw data by fitting a 1D polynomial in each wavelength bin; the spectra 
were then extracted and calibrated in wavelength by applying the inverse dispersion solution (IDS). 
To account for instrument flexure, the IDS was derived from an arc line exposure taken immediately prior 
to or following the scientific one. A second background subtraction pass was then performed on the extracted 
2D spectra, using a background map for each slit constructed by combining individual frames without applying 
any offset. Frames were then offset according to the observing sequence and combined. Each 1D spectrum was 
then extracted using optimal extraction and corrected for the instrument's sensitivity, with a sensitivity 
curve obtained from observations of a red star. 
As the observations were performed during several nights in different meteorological conditions, the 
resulting spectra are in (arbitrary) pseudo-flux units.

\begin{figure}
\centering
\includegraphics[width=0.5\textwidth]{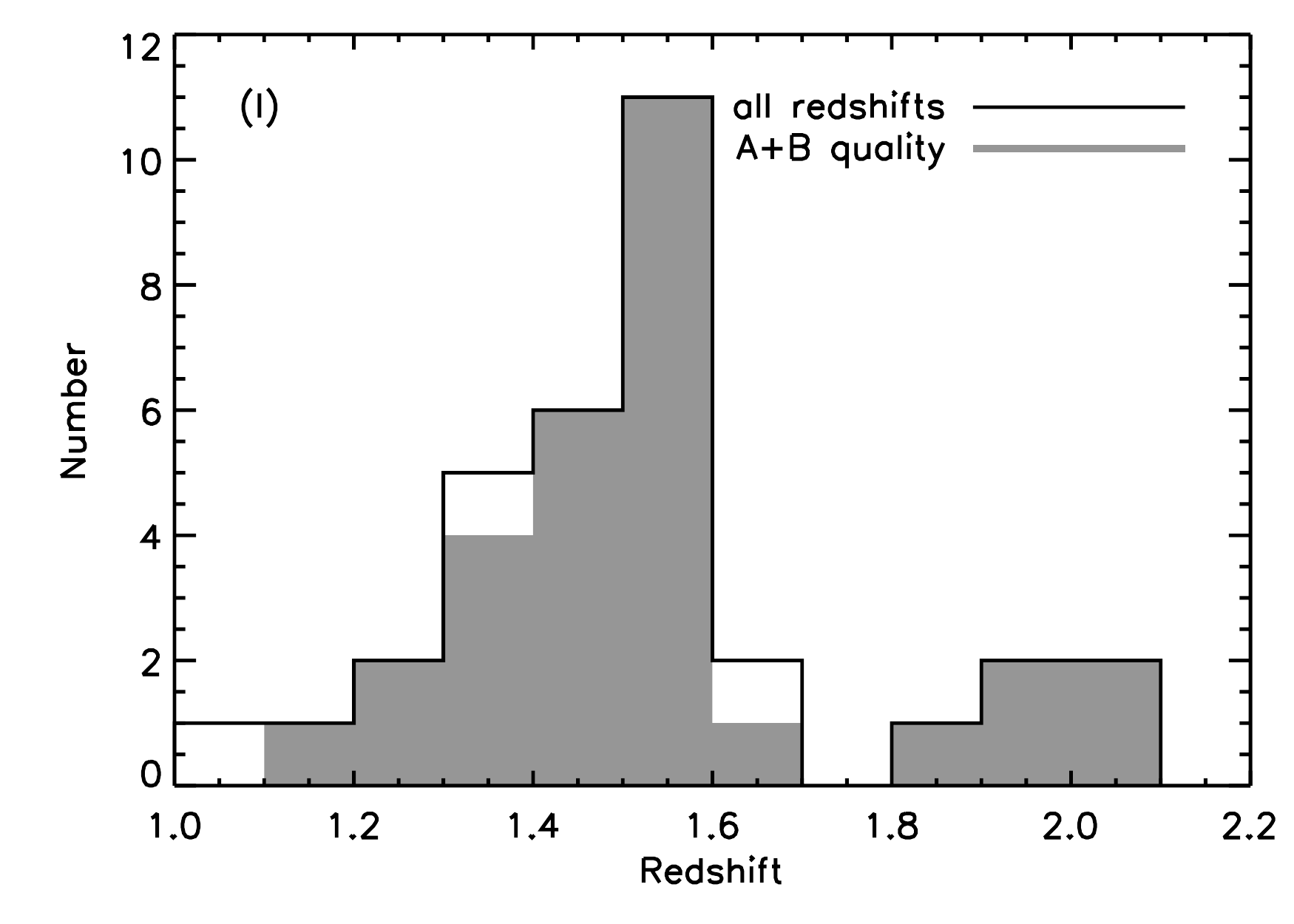}
\includegraphics[width=0.5\textwidth]{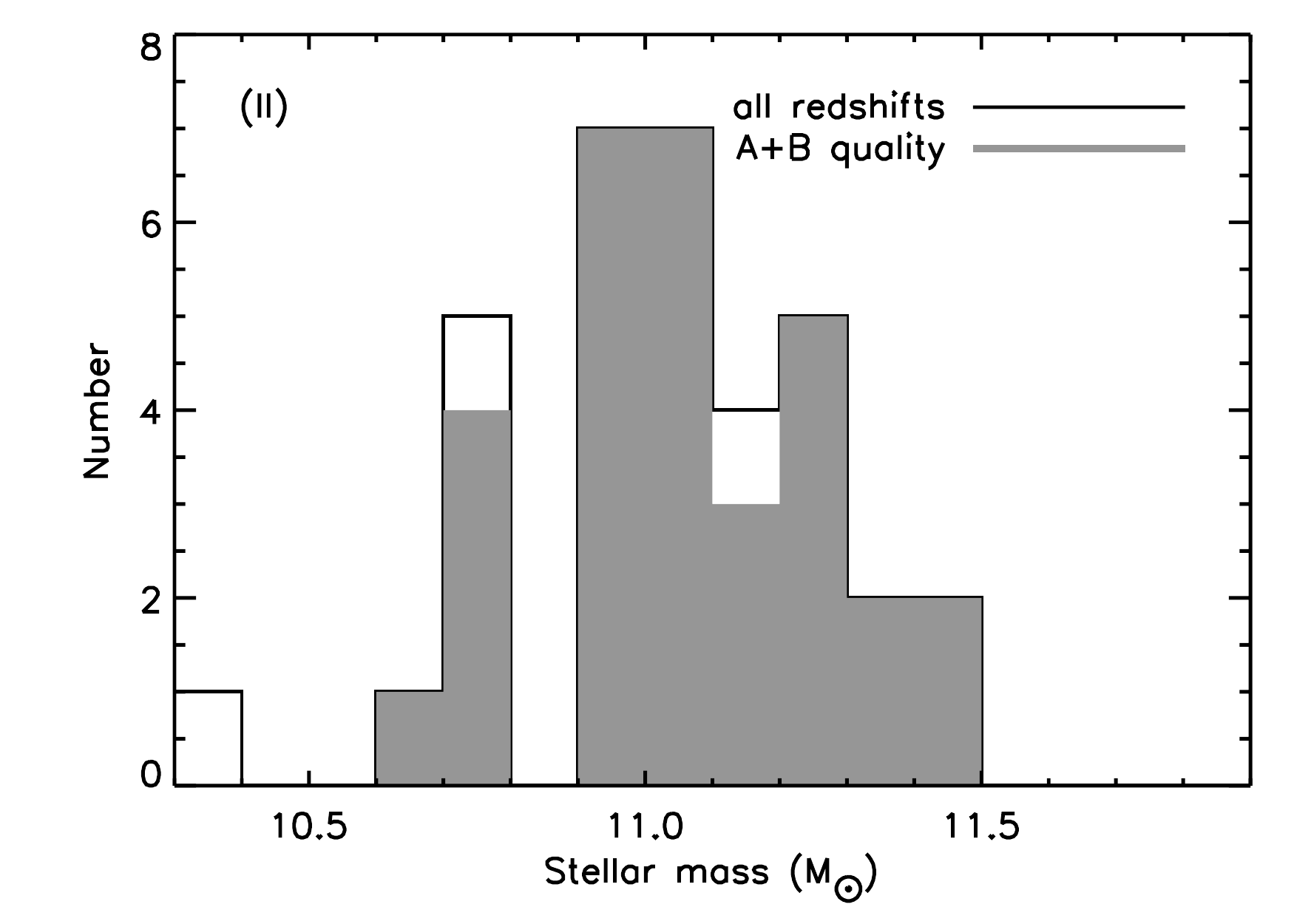}
\caption{\emph{I):} Distribution of spectroscopic redshifts (solid histogram) derived from 
fitting the observed spectra with stellar population models, with the shaded histogram 
showing the secure ones.
\emph{II):} Distribution of stellar masses (Salpeter IMF) derived from 
UltraVISTA photometry, using the same grid of models.
}
\label{fig:hist}
\end{figure}

\subsection{\label{zspec}Redshifts}

\begin{figure}
\centering
\includegraphics[width=0.5\textwidth]{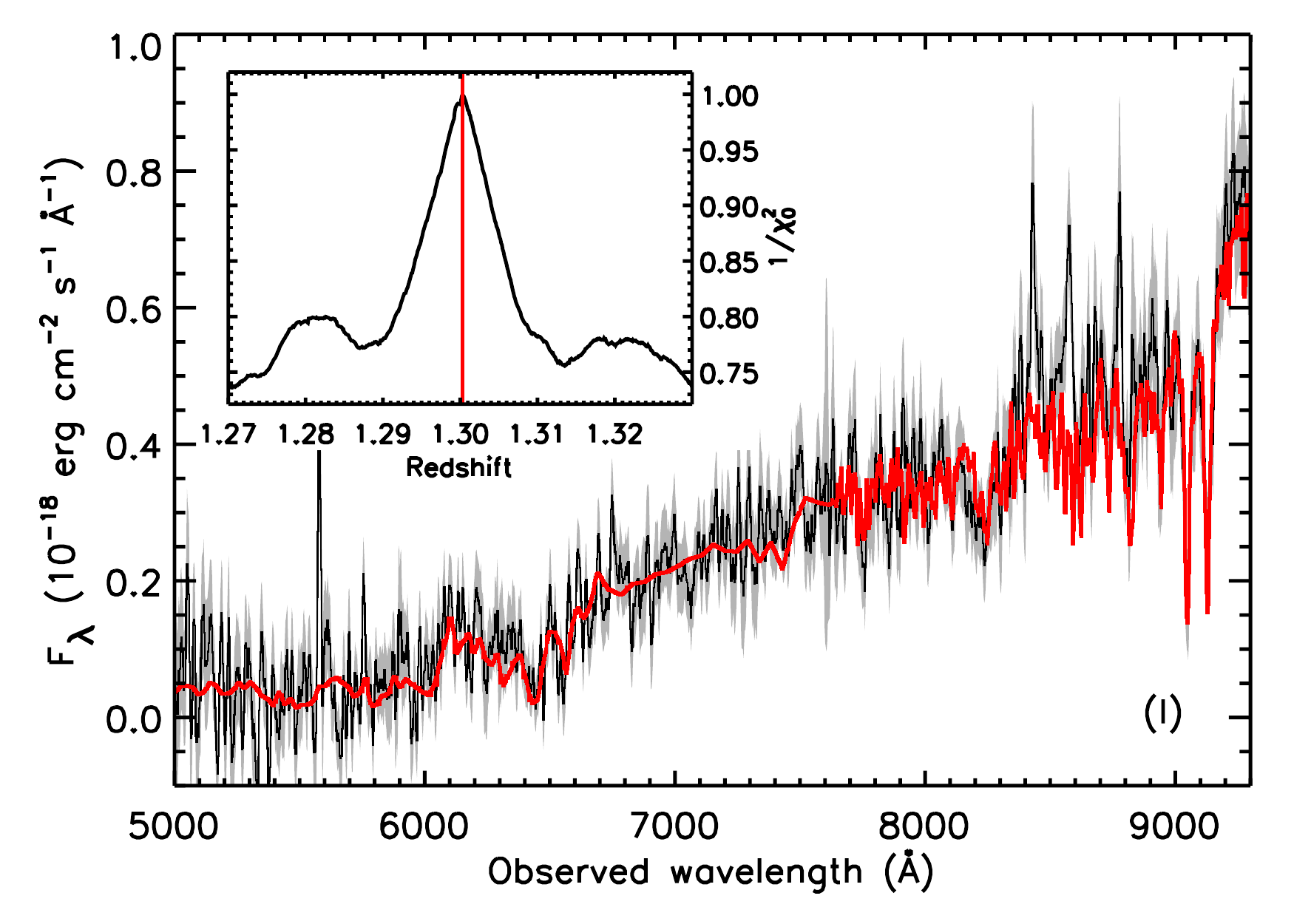}
\includegraphics[width=0.5\textwidth]{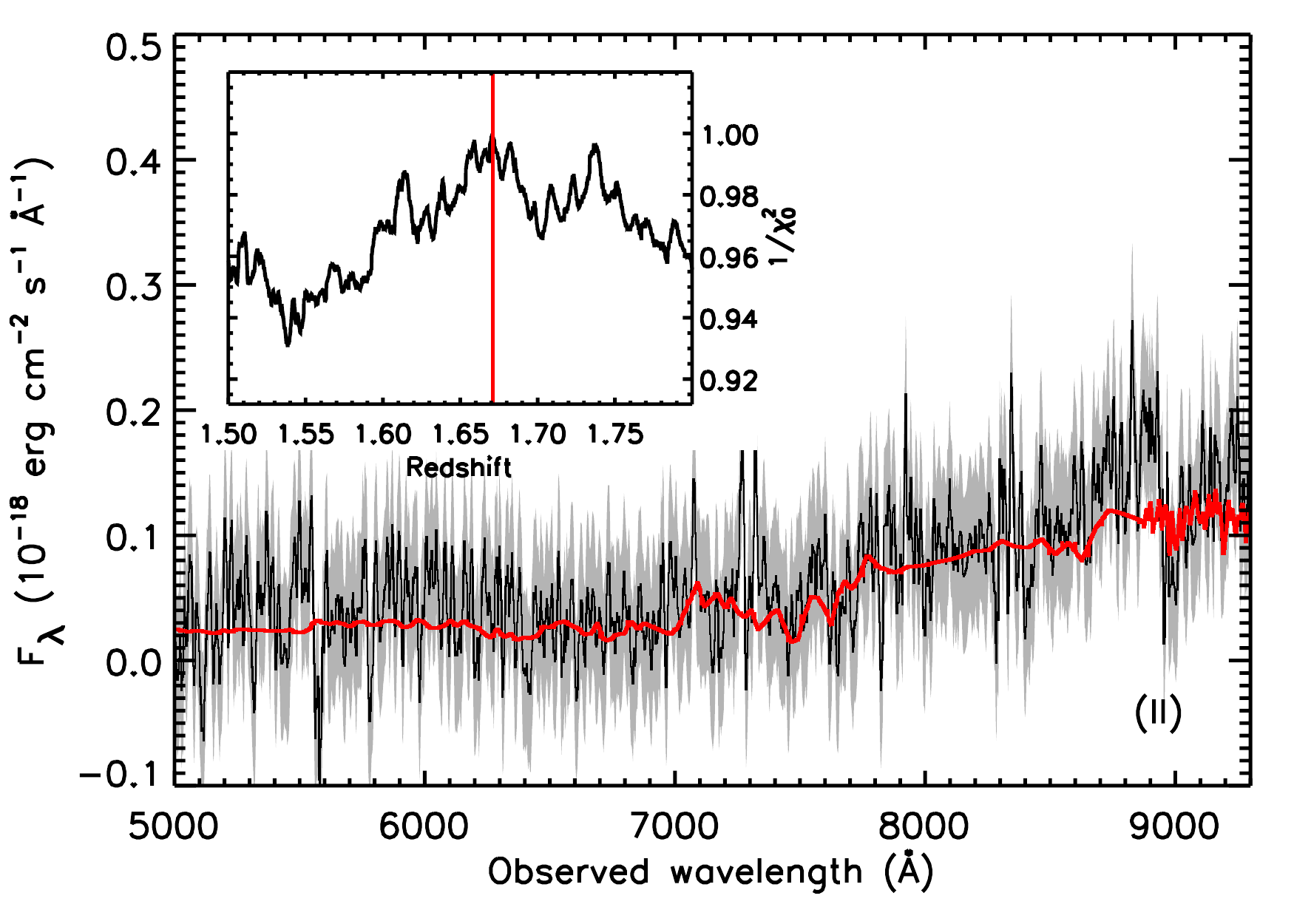}
\caption{Examples of a secure (I; ``A'' quality; ID 225039) and an uncertain (II; ``C'' quality; 
ID 240680) redshift. In each case, the black curve shows the observed spectrum, with error in gray, 
and the red one the best-fit model. The inset shows the inverse of the reduced $\chi^2$, with the 
best-fit solution marked by a red vertical line. Each spectrum has been rebinned to a width of 
5~pixels.
}
\label{fig:zfit}
\end{figure}

We derive redshifts for 34 of the 35 extracted pBzK spectra. The distribution of spectroscopic redshifts, 
which has a median of $z=1.51$, is shown in Fig.~\ref{fig:hist} and the full list given in Appendix~\ref{ztable}. 
We estimate redshifts with two different methods: (i) by measuring absorption lines using the \emph{rvidlines} 
IRAF task and estimating uncertainties from the r.m.s. of centroid positions, or (ii) through fitting of the full 
spectra with templates. 
For the latter method we use either average spectra from the K20 and GMASS surveys \citep{Cim02,Kur13} or a 
grid of stellar population models based on \citet[BC03]{BC03} single stellar population (SSP) templates. In this 
case we assume a delayed, exponentially declining SFH, with ages between 0.1 and $\min(t_z,6)$~Gyr (where $t_z$~is 
the age of the universe at redshift $z$) and characteristic timescale values between 1~Myr and 2~Gyr. We fit 
each composite model (continuum+lines) individually and estimate the confidence interval on the redshift from the 
$\chi^2$~distribution, using for each object the noise spectrum produced by the pipeline as error spectrum.
We include the effect of dust extinction, following the \citet{Cal00} prescription with $E(B$-$V)=0-1$. We adopt 
a regularly spaced redshift grid with an initial coarse step of $\Delta z=0.001$, which we refine to 
$\Delta z=10^{-4}$~once a solution has been identified. 
Because the S/N of the spectra is generally low ($\sim$1~per pixel), this method tends to be more efficient 
at estimating redshifts than simply correlating the positions of (apparent) absorption lines. 
Examples of full-spectrum fits are shown in Fig.~\ref{fig:zfit}. We assign to the resulting redshifts 
a quality flag based on the following criterion:
spectra for which the fit has a unique solution and the two methods yield consistent estimates 
are given the quality flag ``A''; on the other hand, we use the quality flag ``B'' for spectra for 
which the first method fails (i.e., no clear absorption features can be seen in the spectrum) but the 
template fit still yields a single strong solution; finally, we assign quality flag ``C'' to spectra 
for which the fit has one or more secondary solutions and no redshift can be estimated by other methods. 
Redshifts of A or B quality are considered secure, while C quality redshifts are uncertain. 
Uncertainties on the redshifts are estimated from the $\chi^2$~of the template fit, after rescaling so 
that it be 1 per degree of freedom if originally higher. The redshift uncertainties thus derived have a
median value of $\sigma_z=0.0009$.
We obtain 20 A, 11 B, and 3 C quality redshifts, respectively, with only one spectrum in the sample 
having too low S/N for either method to yield a meaningful estimate. Considering only the secure redshifts, 
this corresponds to a success rate of 89\% for the spectroscopically detected objects and a completeness of 
40\% when including all pBzK galaxies in the target area.\\

\begin{figure}
\centering
\includegraphics[width=0.5\textwidth]{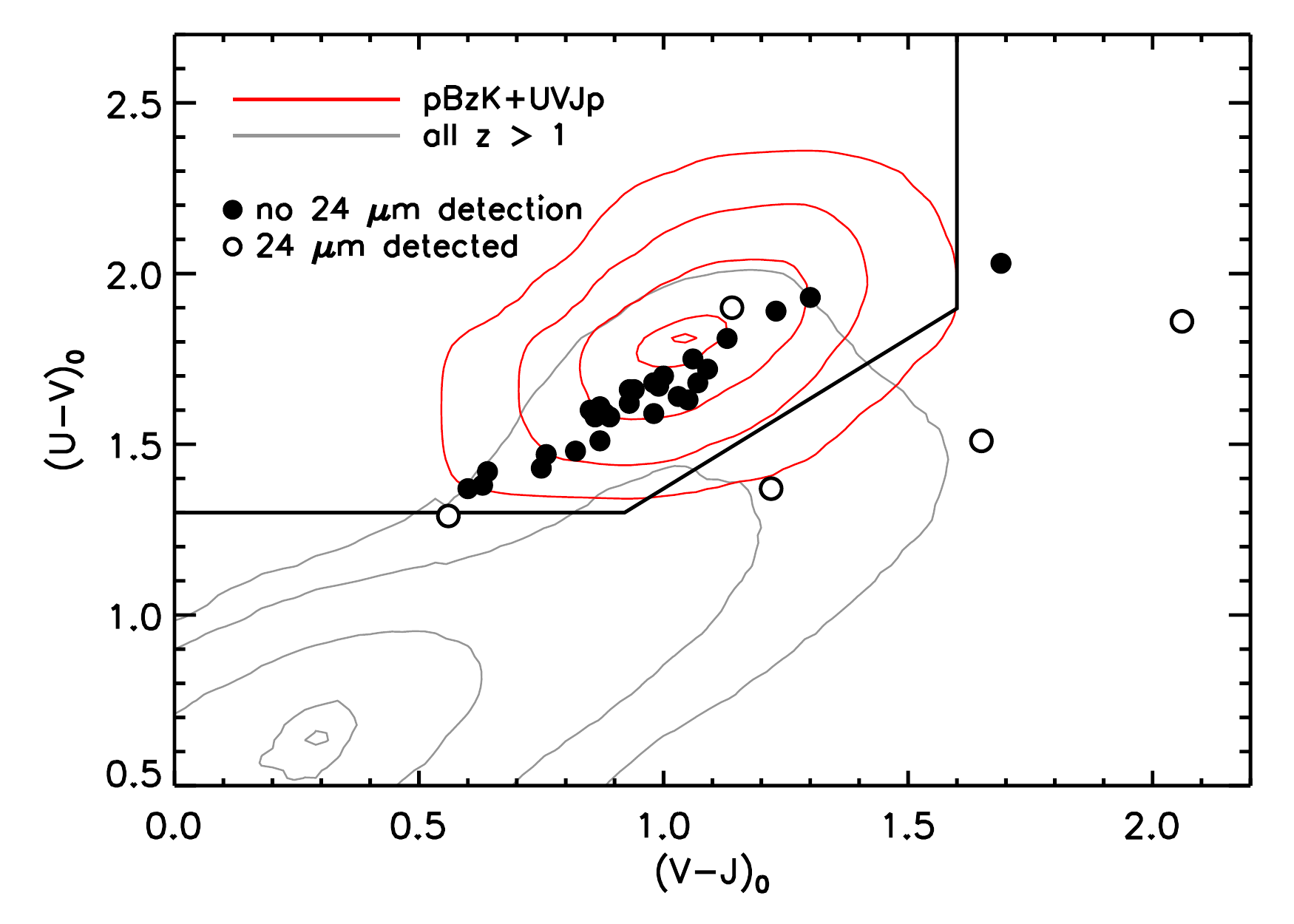}
\caption{Rest-frame $U-V$ vs $V-J$~diagram of galaxies in the COSMOS field. Filled circles mark the 
position of our pBzK galaxy sample in color-color space, with empty ones indicating 24~$\mu$m detections. 
The red and gray contours show, respectively, the distribution of $BzK+UVJ$-selected quiescent 
galaxies (see Sect.~\ref{fir}) and $z>1$ galaxies, using the same relative scale.
}
\label{fig:uvj}
\end{figure}

\begin{figure*}
\centering
\includegraphics[width=0.99\textwidth]{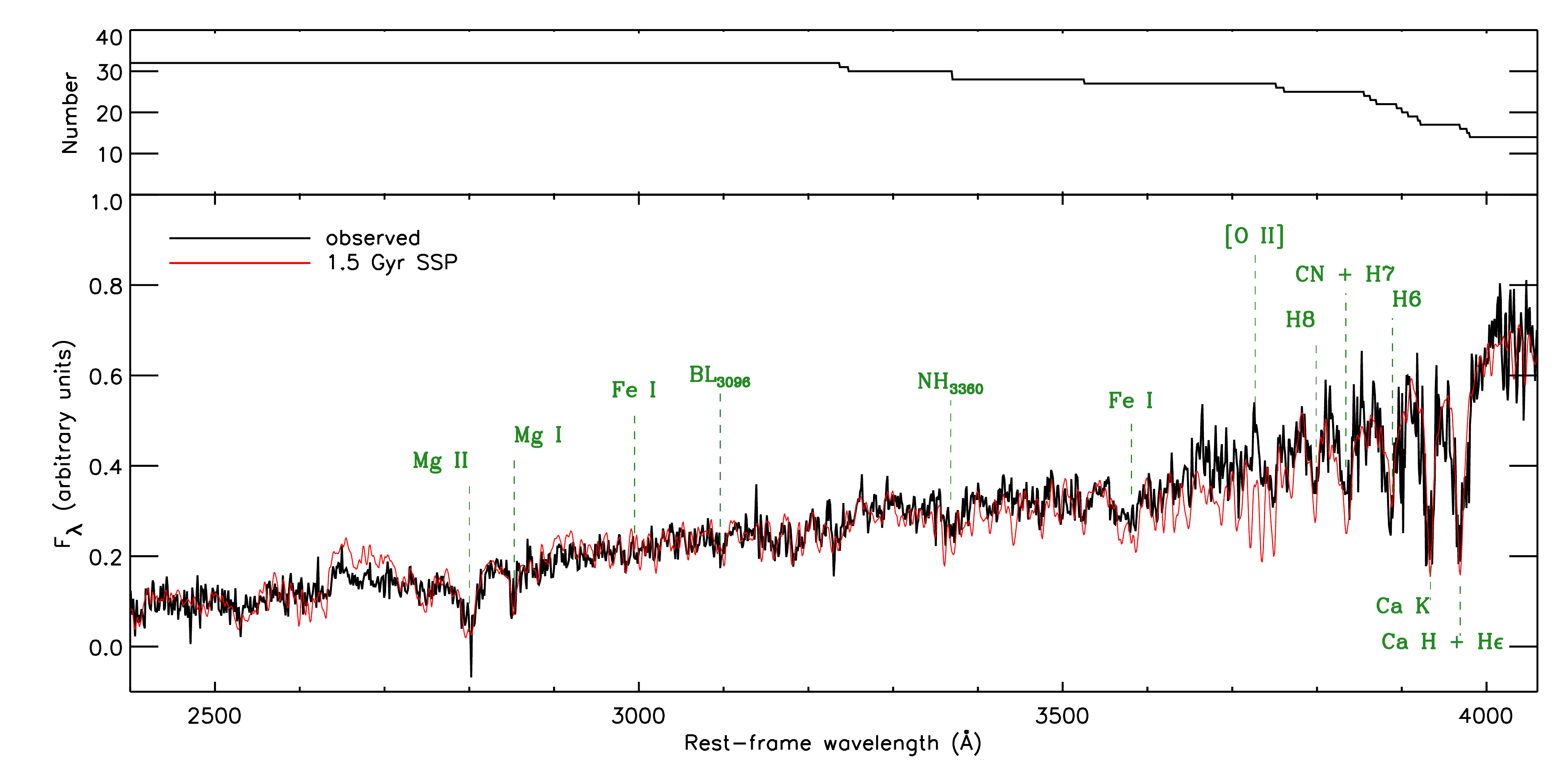}
\caption{\emph{Top:} number of A and B quality spectra contributing to the median spectrum as a 
function of rest-frame wavelength. 
\emph{Bottom:} stack of all A and B quality spectra, as a function of rest-frame wavelength, with a 
1.5~Gyr SSP template \citep{M11} for comparison. Prominent spectral features are indicated in green.
}
\label{fig:spectot}
\end{figure*}

We match the positions of the spectroscopically confirmed pBzKs to the multiband catalog of 
\citet{Muz13}, which is based on the UltraVISTA survey of the COSMOS field \citep{McC12}. 
Ignoring the GALEX bands due to their lower resolution, we model the remaining 28-band photometry 
with the same grid of stellar population models as described above (fixing the redshift but letting 
the other parameters vary) to estimate stellar masses and dust extinction values. The distribution 
of these stellar masses, which has a median of $M_{\star}=1.1\times10^{11}$~M$_{\odot}$, is shown 
in Fig.~\ref{fig:hist}. We note that it is consistent with the BC03 masses given in \citet{Cpo16}, 
although the values for individual galaxies differ owing to differences in the SFH and parameter grid 
used. As a sanity check, we also use the best-fit models to the SED to compute rest-frame $U-V$~and 
$V-J$~colors and confirm that most of the pBzK galaxies are selected as passive using the 
high-redshift \emph{UVJ} criterion of \citet{Wil09}, except for the 24~$\mu$m ones. As shown in 
Fig.~\ref{fig:uvj}, the former are well within the locus of $z>1$~quiescent galaxies.

\section{\label{stelpop}Stellar population properties}

\begin{figure*}
\centering
\includegraphics[width=0.99\textwidth]{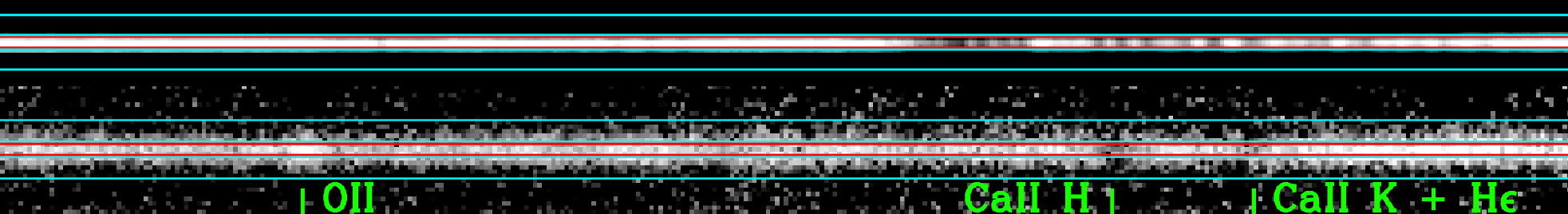}
\includegraphics[width=0.49\textwidth]{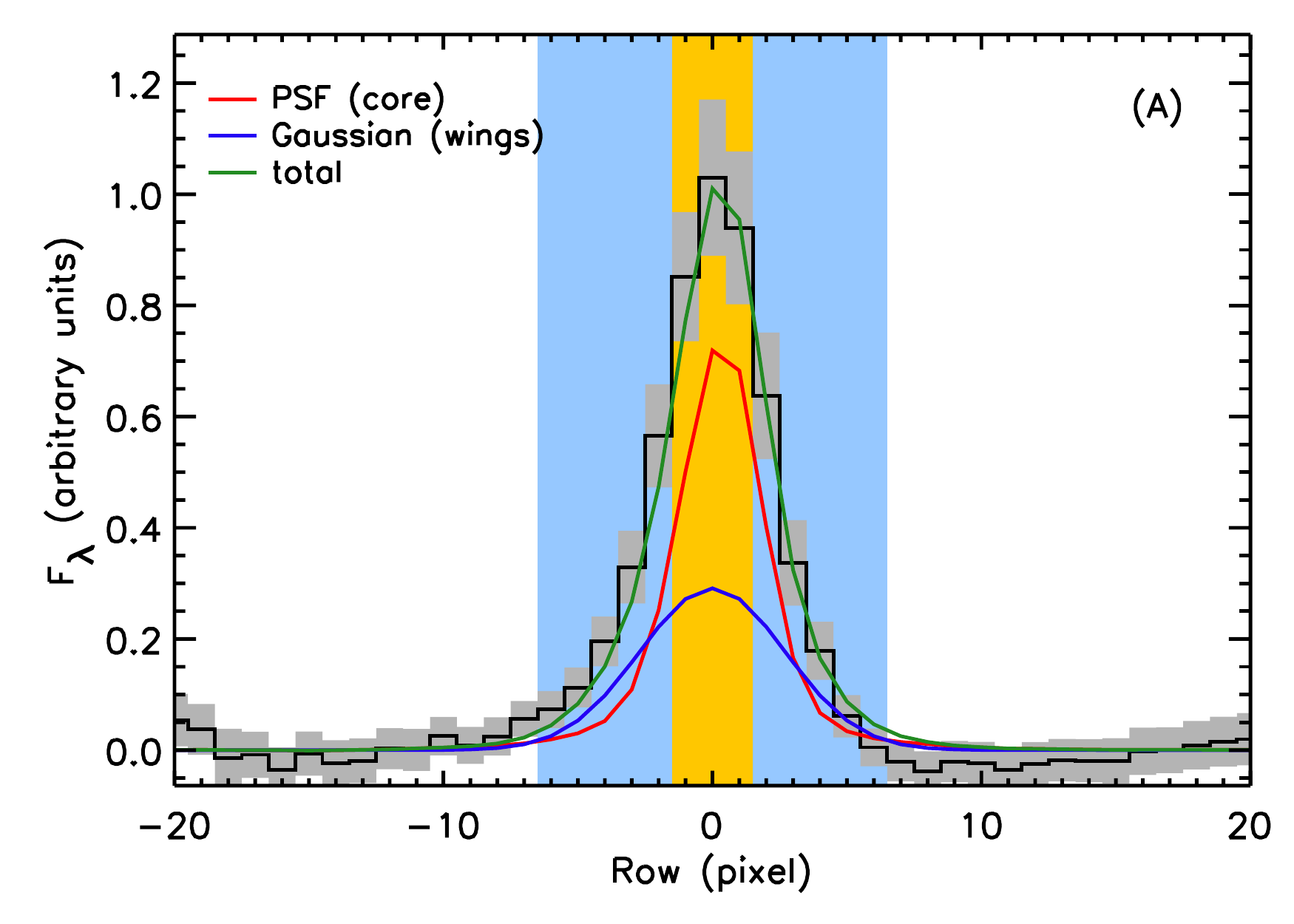}
\includegraphics[width=0.49\textwidth]{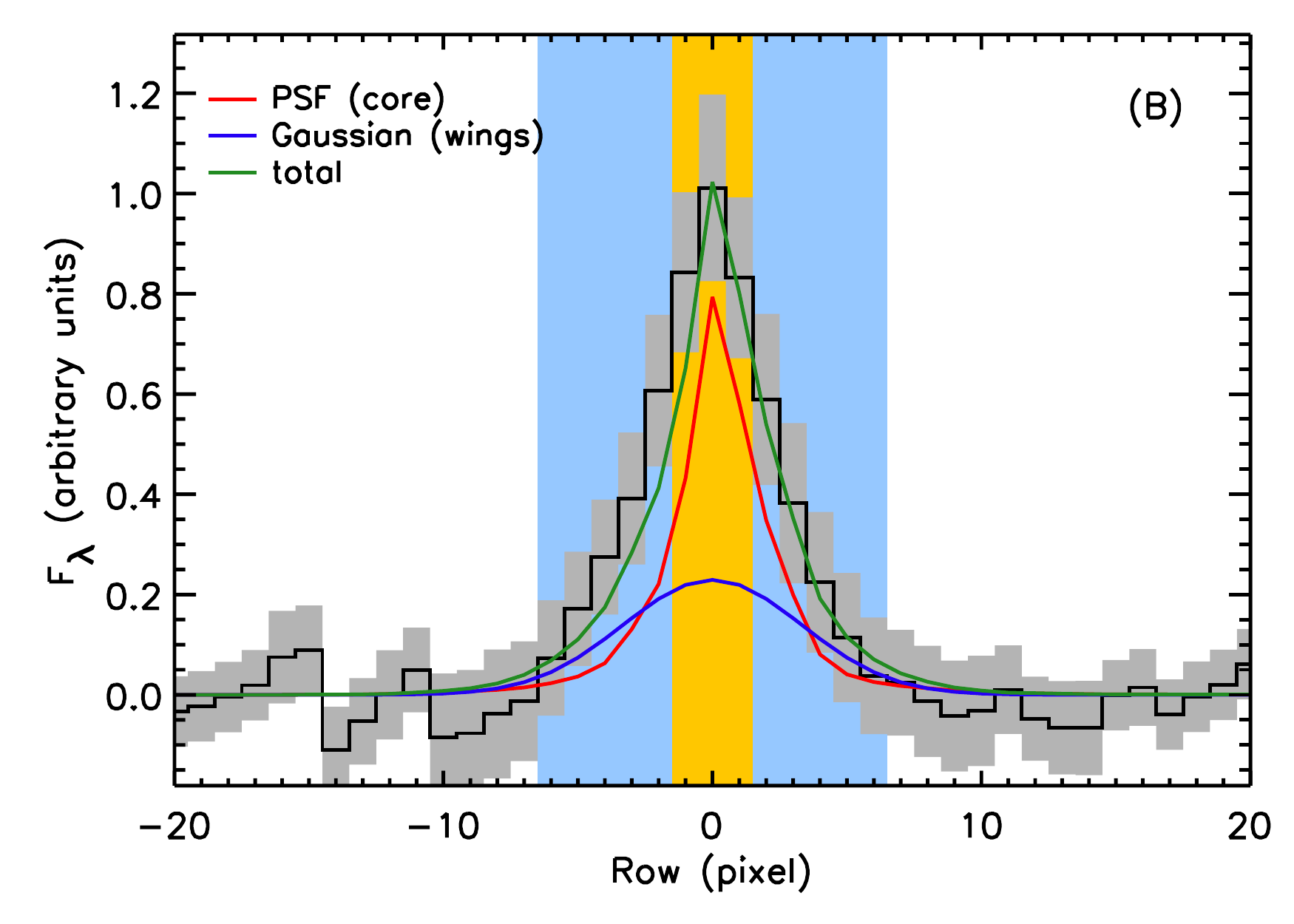}
\caption{\emph{Top:} median 2D spectra of stars (top row) and pBzKs in the range 
$z=1.2-1.5$~(bottom row), in the same wavelength range and using the same relative flux scale. 
Prominent features in the stacked galaxy spectrum are indicated in green. The red and cyan lines 
show the inner and outer extraction apertures, respectively.
\emph{Bottom:} Cross-dispersion profile of the stacked pBzK spectrum at 
$3650-4050$~$\AA$~rest-frame (dark histogram with gray errors), fit with a combination of a 
point-source (red) and a Gaussian (blue) profile. The inner and outer extraction apertures are 
indicated by the orange and light blue shaded regions, respectively. Panels (A) and (B) 
show, respectively, the cross-dispersion profiles before and after correcting for galaxy 
morphology. Only the profile shown in panel (A) was used for the decomposition and analysis 
described in Sect.~\ref{decomp} and below.
}
\label{fig:decomp}
\end{figure*}

Next, we create a median 2D spectrum of the spectroscopically confirmed A and B quality pBzKs, shifting 
each reduced 2D spectrum to the median redshift of $z=1.51$~and correcting for cosmological dimming. 
We then shift the 2D spectra in the spatial direction so that the peak of the spectral trace 
falls onto the central pixel. For each source, we estimate the $y$ position of the trace by fitting 
a simple Gaussian peak to its median profile along the dispersion direction. We find that the 
r.m.s. offset of individual spectra in the spatial direction is $\lesssim0.3$. 
Individual spectra are then here weighted by their intrinsic rest-frame UV-optical luminosity. 
We also compute a 2D variance spectrum, where the value of each pixel corresponds to the median 
absolute deviation of fluxes at each pixel position in the stack. 
We note that the typical redshift uncertainty corresponds to $<$50\% of the spectral resolution, and 
that it thus has negligible impact on the stack. The resulting 1D spectrum, extracted over the entire 
cross-dispersion column, is shown in Fig.~\ref{fig:spectot} and will be made available as an ASCII table 
at the CDS (the file contains the following information: Column 1 gives the rest-frame wavelength in 
$\AA$, while Column 2 and 3 give, respectively, the flux and flux uncertainty estimated from the 
variance spectrum, both in arbitrary $f_{\lambda}$ units). 
We characterize this spectrum using SSP templates from \citet[hereafter M11]{M11}, with resolution 
matched to that of our VIMOS spectra. These templates are based on the MILES stellar library \citep{SB06} 
but have been extended to the far UV and thus cover the whole rest-frame wavelength range spanned by 
the stacked spectrum. They have ages between 0.1 and 4~Gyr and four different metallicities of 2\% to 
twice the solar value. We estimate uncertainties on derived parameters from the 68\% confidence 
intervals of the $\chi^2$~fit. We note that flux errors derived from the variance spectrum are 
a factor $\sim$10 higher than the r.m.s. noise in the background regions of the 2D median spectrum, 
yielding conservative uncertainty estimates on the derived parameters. We find a luminosity-weighted 
age of $1.5^{+1.5}_{-0.5}$~Gyr, and use the high-S/N rest-frame UV absorption features 
\citep[Mg II, Mg I, NH 3360, and BL 3580;][]{Fan90,Dav94,Ser11} to constrain the luminosity-weighted 
metallicity at $1^{+0.6}_{-0.2}$~$Z_{\odot}$. These values are consistent with the range of derived 
ages and metallicities in massive quiescent galaxies at this redshift, from either rest-frame UV or 
rest-frame optical spectra \citep{Cim04,MCy04,Dad05,Ono15}.\\ 

This median spectrum also displays two other noteworthy features: 
First, it shows clear [OII]3727$\AA$~line emission. Second, it is marginally resolved, as its profile 
in the cross-dispersion (spatial) direction is more extended than the stacked spectrum of 3 stars taken 
during the same observation (Fig.~\ref{fig:decomp} and \ref{fig:stelpop}).
To investigate this further, we redo the stack considering only the pBzKs with redshift below the median 
($z<1.5$), i.e., whose spectrum is not truncated before the 4000~$\AA$ break, since it is the region with 
the most useful features for stellar population modeling. This also has the advantage 
that the resulting subsample covers $\sim$1~Gyr of cosmic time and thus a relatively narrow epoch in 
the galaxies' evolution.
We also remove possible AGNs from the stack by selecting only sources which are not detected in 
\emph{Spitzer}/MIPS 24~$\mu$m data and not classified as AGNs either in the photometric catalog or 
with the updated IRAC color criterion of \citet{Don12}.
This criterion yields 13 sources with a median redshift of $z=1.4$ and a median stellar mass of 
$M_{\star}=1.1\times10^{11}$~M$_{\odot}$, identical to that of the parent sample. A portion of the 
resulting median stack is shown in Fig.~\ref{fig:decomp}. We estimate uncertainties for each pixel 
using bootstrap resamplings of the data, with sizes of half the initial sample.\\

\subsection{\label{decomp}Spatial decomposition and spectral extraction}

We then extract 1D spectra from the 2D composite spectrum in three different apertures: an inner 
3-pixel one centered on the peak of the cross-dispersion profile, an outer aperture extending 5 
pixels on each side of the central aperture (Fig.~\ref{fig:decomp}), and a full (inner + outer) 
one. These widths were chosen so that the inner and outer extracted spectra are mostly independent 
from each other, the width of the inner aperture being comparable to the FWHM of the stellar stack 
(3.5~pixels), while also minimizing differences in S/N between the two spectra. 
As shown in Fig.~\ref{fig:wings}, they are broadly similar except around 4000~$\AA$, 
where the outer aperture spectrum appears to have a shallower 4000~$\AA$ break (we note however 
that there is not enough continuum redward of the break to compute a spectral index) and a deeper 
3969~$\AA$~line than the inner one, suggesting excess H$\epsilon$~absorption and thus more recent 
star formation.\\

\begin{figure}
\centering
\includegraphics[width=0.5\textwidth]{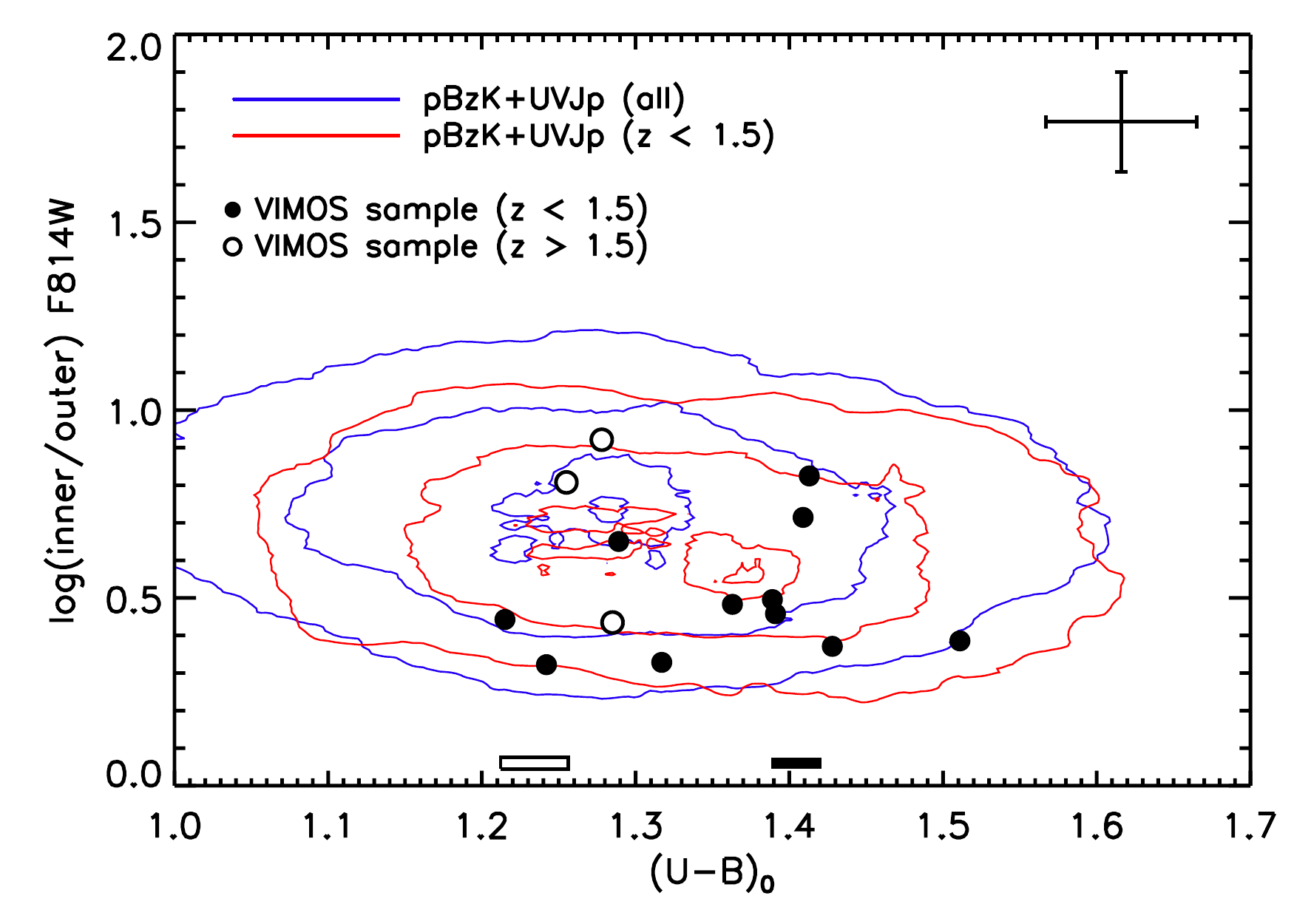}
\caption{Ratio of rest-frame UV flux in the inner and outer aperture, based on the \emph{HST}/ACS F814W 
mosaic, as a function of rest-frame $U-B$~color. Filled and empty circles show pBzKs below and above 
$z=1.5$, respectively, using the same relative scale. The blue and red contours mark the distribution of 
$BzK+UVJ$-selected galaxies (see Sect.~\ref{fir}) in the full $z=1.5-2.5$~range and at $z<1.5$, respectively. 
The error bars in the upper right corner show the typical uncertainty on both quantities while the filled 
and empty bars show the rest-frame color distribution of low-S/N pBzKs at $z<1.5$ and $z>1.5$, respectively.
}
\label{fig:acs}
\end{figure}

This apparent stellar population difference could be due to our including in the stack populations 
of galaxies quenched at different epochs, with the younger ones being also larger and thus 
contributing more to the flux in the outer aperture \citep[see, e.g.,][]{Car13,Wll16}. 
To test this we look at the publicly available \emph{HST}/ACS F814W mosaic \citep{Koe07}, corresponding 
here to the rest-frame $3300-3700\AA$. We measure the ratio of F814W flux in two apertures of diameter 
equal to the spectroscopic ones (a measure of compactness in the rest-frame UV) and compare it to the 
rest-frame $U-B$~color of the galaxies (which measures the amplitude of the 4000~$\AA$ break). Only 14 
out of 34 spectroscopically galaxies have high enough S/N in the F814W image to allow for such a  
comparison. Since we only consider galaxies with high enough surface brightness, this could effectively 
be assimilated to a selection in age. However, the rest-frame colors of low-S/N pBzKs are consistent 
with those of the higher-S/N galaxies. We perform this test for both these 14 pBzKs and a sample of all 
$BzK$- and $UVJ$-selected galaxies in the COSMOS field (see Sect.~\ref{fir} for details on this second 
selection). As shown in Fig.~\ref{fig:acs}, we find no correlation between the ratio of F814W light in 
the inner to outer aperture, measured using the circularized profile of the detected pBzKs, and their 
rest-frame $U-B$~color estimated from the SED fit. This suggests that compactness in the rest-frame UV 
is not strongly correlated with the combination of age and metallicity traced by the color. Both the 
color and compactness of our pBzKs also appear to be consistent with the distribution of the larger 
sample.
While we see no evidence for a correlation between the global stellar populations of our galaxies 
and their compactness, the combination of differently-sized objects can still have an effect on derived 
properties in the inner and outer apertures. Assuming that all galaxies in our sample have similar 
stellar population structure, we estimate from Fig.~\ref{fig:acs} that their combination in the stack 
should make us underestimate stellar population differences between the two apertures by $\sim$10\%. 
This is however well within the uncertainties we derive below.\\ 

If we then assume that the stellar populations of the pBzKs can be spatially separated into two 
components, one compact and central (``core'') and the other extended (``wings''), the intrinsic spectra 
of the core and wings components ($f_{\text{c}}$~and $f_{\text{w}}$, respectively) are related to the 
observed spectra in the inner and outer apertures ($f_{\text{i}}$~and $f_{\text{o}}$, respectively) as 
follows:

\begin{equation}\label{eq:dec}
\left(\begin{array}{c}
f_{\text{i}} \\
f_{\text{o}} \end{array} \right) = \left(\begin{array}{cc}
w_{\text{ic}} & w_{\text{iw}} \\
w_{\text{oc}} & w_{\text{ow}} \end{array} \right) \left(\begin{array}{c}
f_{\text{c}} \\
f_{\text{w}} \end{array} \right) + \left(\begin{array}{c} 
\delta f_{\text{i}} \\
\delta f_{\text{o}} \end{array} \right),
\end{equation}

\noindent
where $\delta f$~is the noise per pixel column in each aperture and the mixing parameters $w$~are 
the fraction of light in either aperture coming from the core and wings spectra.
We consider two different decompositions of the spectrum: in the first, we parameterize the cross-dispersion 
profile of the stacked pBzK spectrum as a combination of a point source (for the core) and a Gaussian (for 
the wings), as shown in Fig.~\ref{fig:decomp}. We use the profile from a 2D stack of 3 stellar spectra for 
the point source and perform the decomposition on the median profile stacked along the dispersion direction; 
under these assumptions, 
$\left[w_{\text{ic}},w_{\text{iw}},w_{\text{oc}},w_{\text{ow}}\right]=\left[0.76,0.24,0.48,0.52\right]$, 
with uncertainties of $\left[0.03,0.03,0.02,0.02\right]$ (we note that these contribute negligibly to 
the error spectrum, compared to the pixel variance). The resulting core and wings spectra are 
shown in Fig.~\ref{fig:wings}. For the second, based on the observed $i$-band morphologies of massive 
early-type galaxies (ETGs) in the COSMOS field \citep{Mni10}, we assume that the tridimensional distribution 
of stars in the pBzK galaxies can be described by a deprojected de Vaucouleurs density profile \citep{MM87}. 
We then define the core (respectively, wings) as the light within (outside) 4.6~kpc, the effective radius 
estimated from fitting a 1D de Vaucouleurs profile to the observed cross-dispersion profile 
\citep[we note that this value is consistent with the ACS sample of][]{Mni10}.
We reproject and convolve these two fractional profiles with the stellar point spread function to find their 
respective contributions to the inner and outer apertures. This yields 
$\left[w_{\text{ic}},w_{\text{iw}},w_{\text{oc}},w_{\text{ow}}\right]=\left[0.61,0.39,0.44,0.56\right]
\pm\left[0.002,0.002,0.001,0.001\right]$ and 1D spectra similar to the ones obtained with the first 
parameterization.\\

We then perform a series of tests to check the robustness of our profile decomposition. First, we 
shift the median spectrum by $\pm$0.5~pixel in the spatial direction, to conservatively simulate the 
uncertainty in the trace position. We find that this changes the mixing parameters by 0.01 for the core 
and 0.015 for the wings. Next, to probe the effect of combining differently-sized sources on the median 
profile, we construct a 2D median spectrum where the individual cutouts have been corrected for galaxy 
sizes. We rescale the 2D spectra of galaxies detected in the F814W mosaic by the ratio of their 
size in the slit (estimated from their circularized effective radius, axis ratio, and position angle) to 
the median value, while leaving those of undetected sources untouched. The expanded or contracted 2D 
spectra are then drizzled back onto the original grid, assuming empty pixels for missing data. Since this 
effectively modifies the spatial resolution of the spectrum, we repeat this operation using, for each 
galaxy, the 2D stellar spectrum to derive a modified point source profile. This increases 
$w_{\text{ic}}$~and decreases $w_{\text{iw}}$~by 0.1. On the other hand, the size-corrected 2D spectrum 
samples the stacked profile less well, as objects larger than the median are rebinned to a coarser 
pixel grid. This procedure also yields higher and more correlated per-pixel noise, resulting in higher 
flux uncertainties on the extracted 1D spectra. Consequently, quantities derived from the analysis 
presented in Sect.~\ref{specfit} become unconstrained when using the size-corrected spectra, although 
they appear to follow the same trends. Finally, we note that all these different mixing parameters yield 
extracted core and wing spectra that are consistent with each other within uncertainties.

\subsection{\label{specfit}Spectral modeling}

\begin{figure}
\centering
\includegraphics[width=0.5\textwidth]{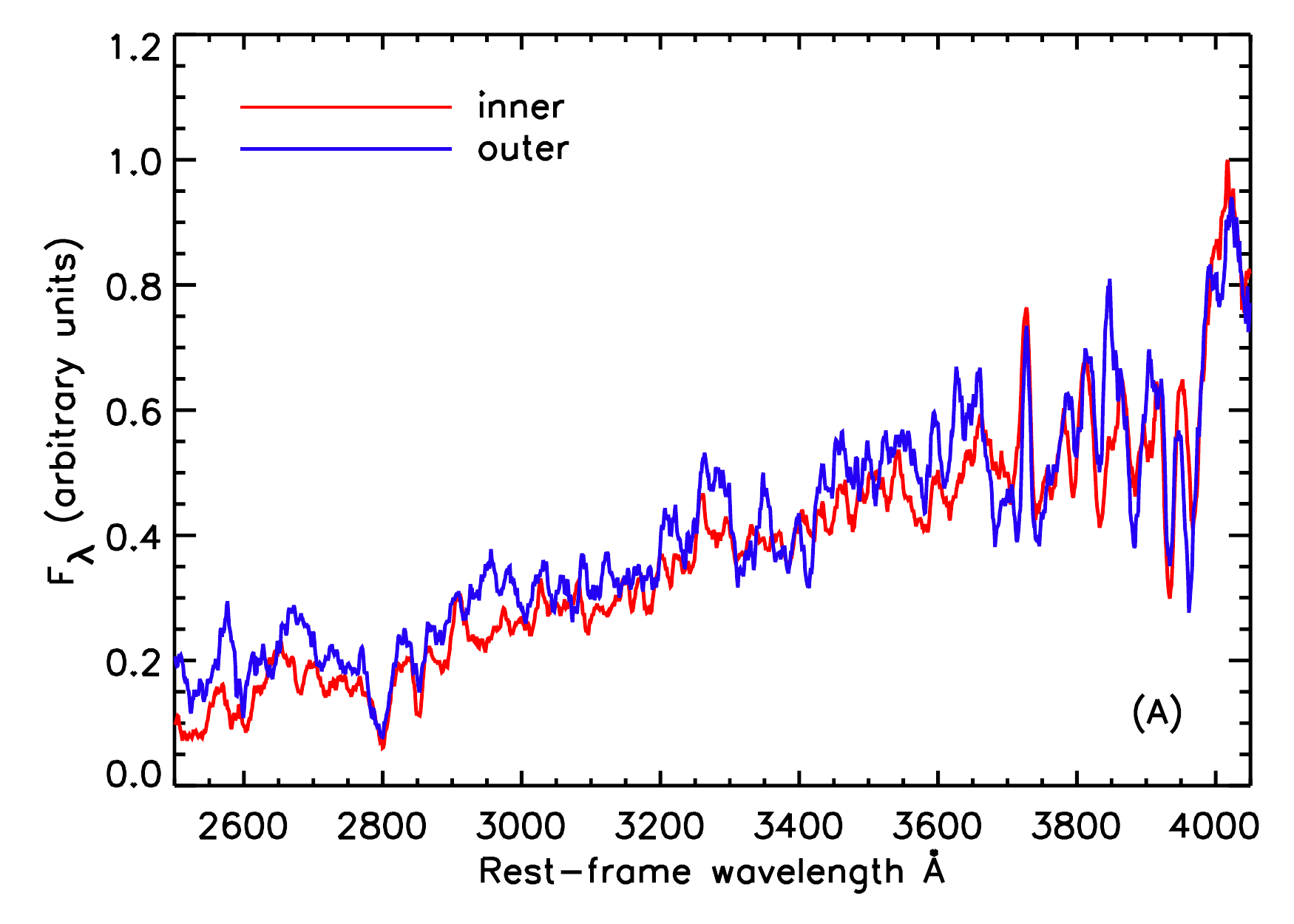}
\includegraphics[width=0.5\textwidth]{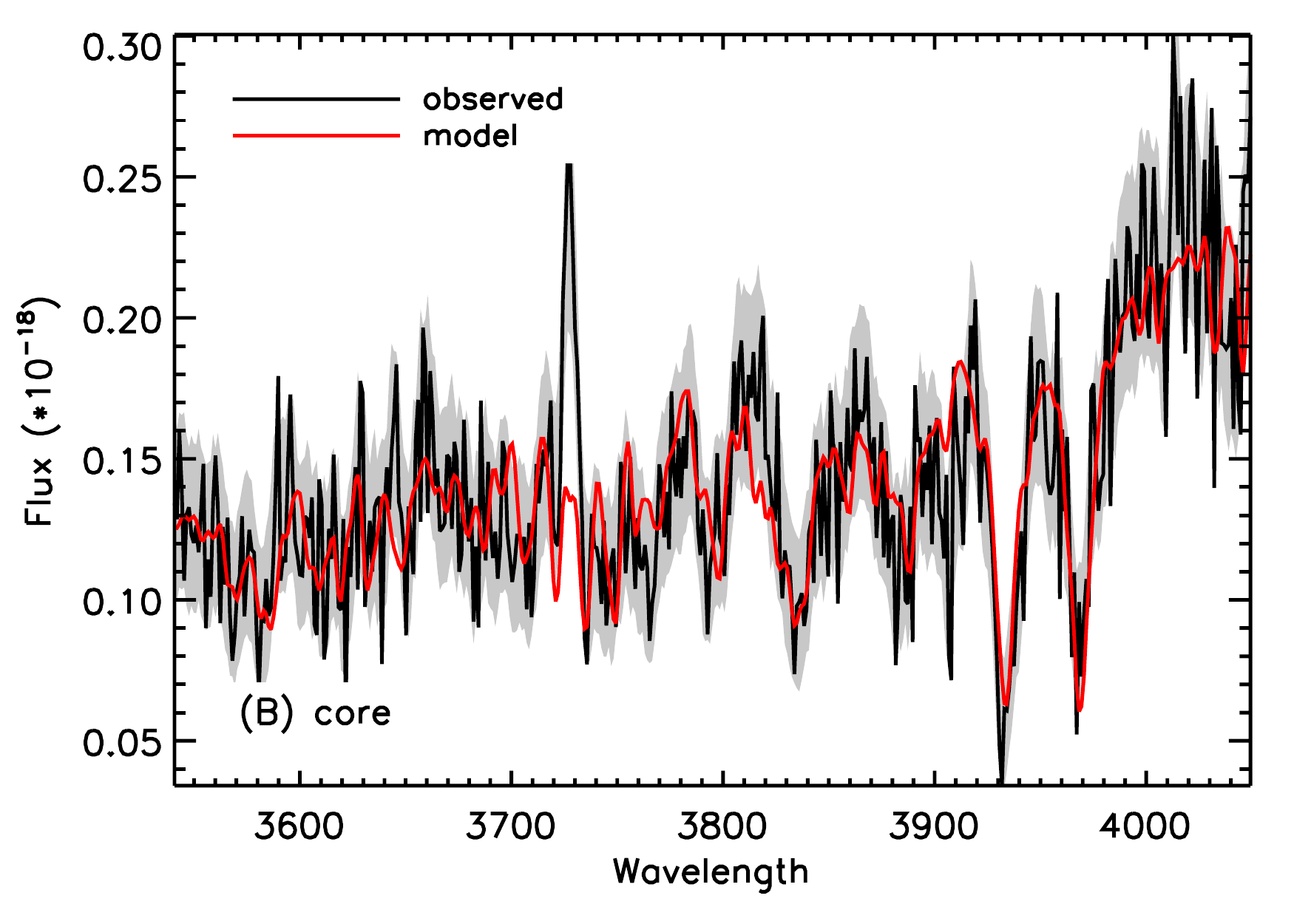}
\includegraphics[width=0.5\textwidth]{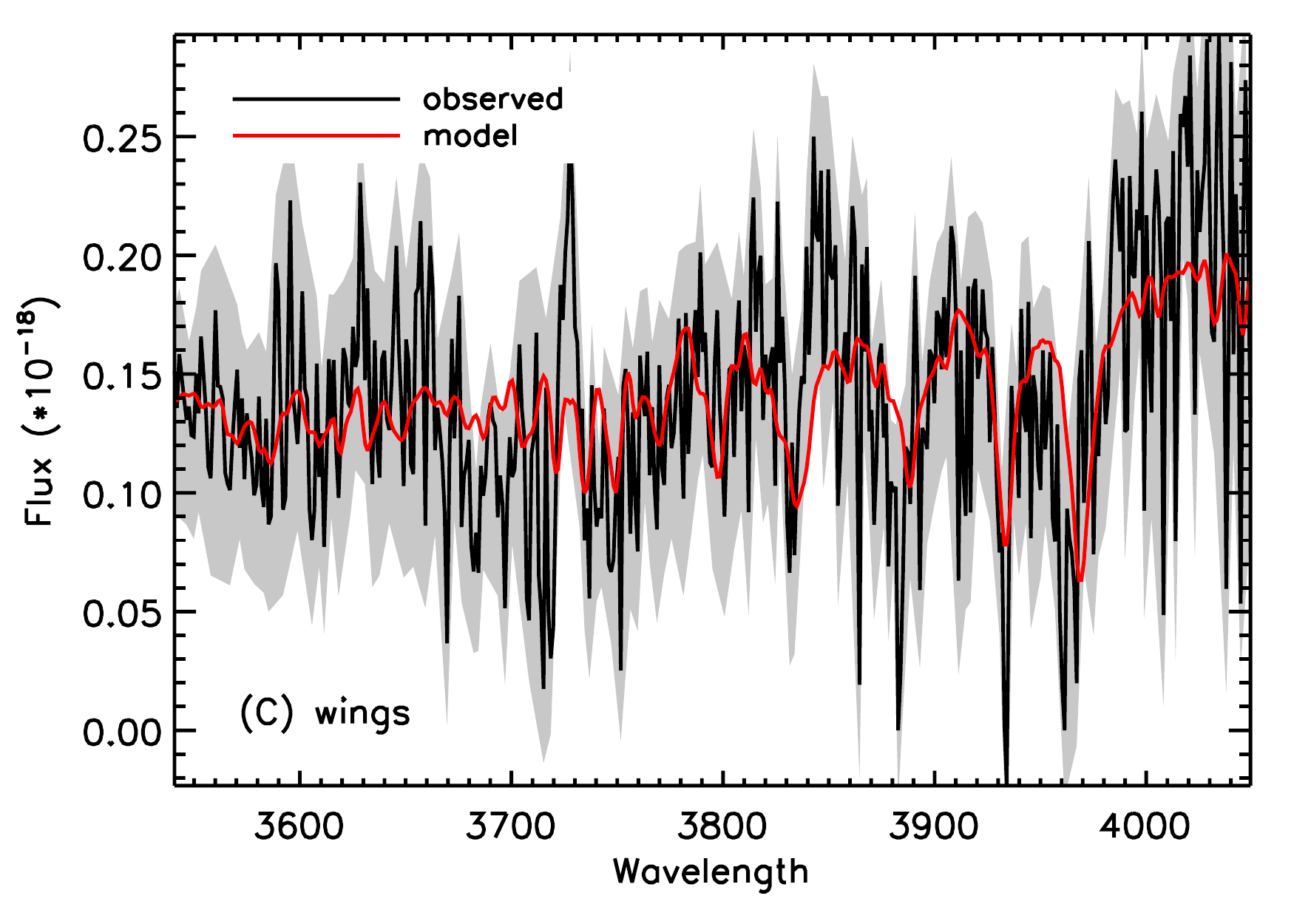}
\caption{\emph{A):} Extracted inner (red) and outer (blue) spectra rebinned to a 10~pixel width. 
\emph{B) and C):} unrebinned core and wings spectra, respectively, decomposed according to 
Eq.~\ref{eq:dec} (black), with associated uncertainties (gray) and best-fit combination of SSP 
templates (red).
}
\label{fig:wings}
\end{figure}

\begin{figure}
\centering
\includegraphics[width=0.5\textwidth]{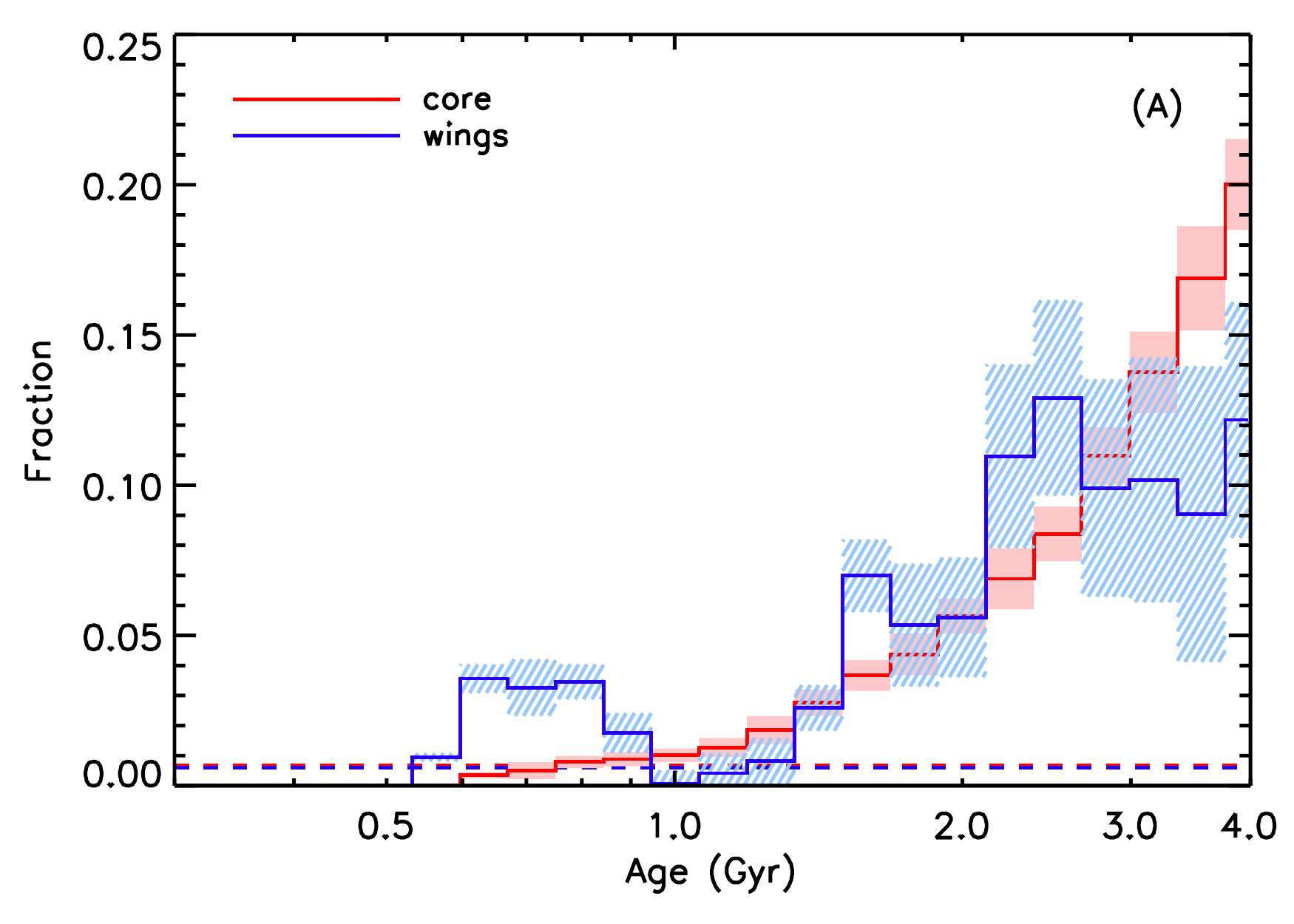}
\includegraphics[width=0.5\textwidth]{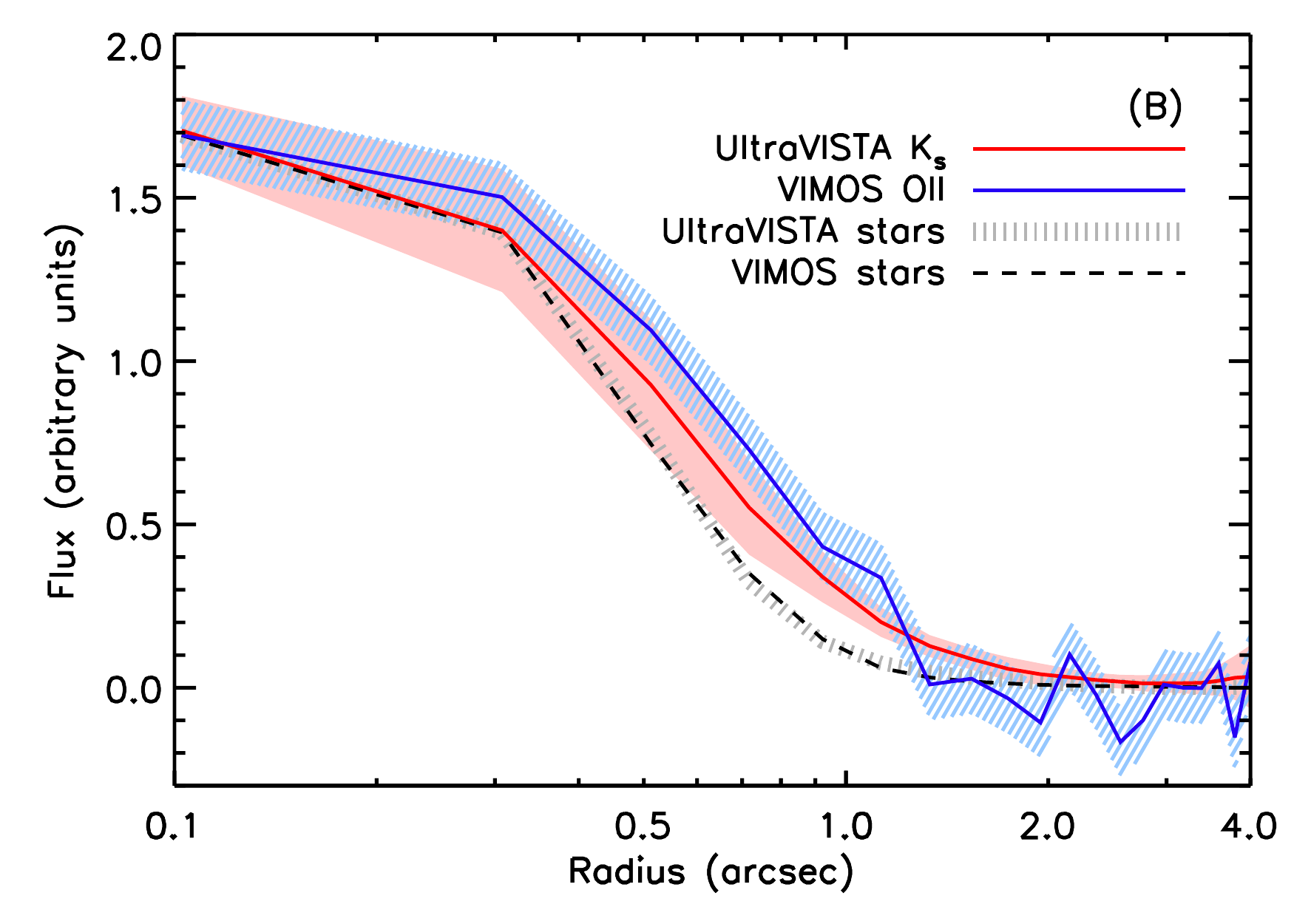}
\caption{
\emph{A):} Relative weight as a function of age of V10 SSP templates in the fit to the 
core and wings spectra (red and blue, respectively). The horizontal dashed lines show the 
level of the SFR derived in Sect.~\ref{o2}, if we assume that the red and blue histograms 
represent the SFH of the core and wings, respectively. 
\emph{B):} Radial profiles of the [OII]3727$\AA$~line from the stacked VIMOS spectrum 
(blue) and $K_s$-band light (red) from UltraVISTA imaging, for the same galaxies. 
Both profiles are shown in 0.205'' bins and have been normalized to the same flux. 
The dashed and dotted curves show the profile of stars in our VIMOS spectra and 
the $K_s$-band image, respectively.
}
\label{fig:stelpop}
\end{figure}

We model the extracted spectra with linear combinations of high-resolution SSP templates. In addition 
to the aforementioned M11 models, we also consider those of \citet[hereafter V10]{Vaz10}, also 
based on the MILES library but computed with different stellar evolution prescriptions. They also 
cover a larger metallicity range, with $\log(Z/Z_{\odot})=-2.32~\text{to}~0.22$, but are only 
defined down to 3540~$\AA$. As previously stated, we consider ages between 0.1 and 4~Gyr. 
We convolve the templates with a fixed Gaussian to match the resolution of the spectra and fit for the 
templates weights with MPFIT \citep{Mar09}, using bounds to enforce nonnegativity. We do not 
attempt to fit the kinematics of this sample since the spectral resolution, which at this redshift 
corresponds to $\gtrsim$200~km/s, is of the order of the expected velocity dispersion. We use the 
1$\sigma$~errors on the template weights, rescaled so that the $\chi^2$~is one per degree of freedom, 
to estimate the uncertainties on the quantities derived from the fit. We restrict ourselves to 
$<$4050~$\AA$~rest-frame, since the flux calibration becomes unreliable at redder wavelengths. We also 
mask the region of the spectrum corresponding to the [OII]3727$\AA$~line. The best-fit models to the core 
and wings spectra are shown in Fig.~\ref{fig:wings} and the reconstituted SFH (i.e., the distribution of 
template weights as a function of age) for both core and wings spectra is shown in Fig.~\ref{fig:stelpop}.\\

Since the core spectrum has higher S/N than the wings, we also perform a simulation to test 
the dependency of the difference between their best-fit solutions to the S/N. We use the best-fit model 
to the core spectrum, adding noise to match the S/N of the wings spectrum, and fit it to the same grid 
of templates. We find that the best-fit SFHs to the core and wings spectra are consistent with 
one another for ages $>$1~Gyr, but that the best-fit wings model includes a significantly higher fraction 
of stellar populations with ages $<$1~Gyr than the best-fit core model, suggesting more protracted star 
formation in the galaxies' outskirts. The relative mass fractions $f_{<1}$~of such young stellar populations 
are $f_{<1}=3\pm1$\% ($4\pm1$\%) in the core and $f_{<1}=13\pm1$\% ($21\pm7$\%) in the wings with the V10 
(M11) templates and for the Gaussian decomposition. In the case of a de Vaucouleurs profile, these values 
become, respectively, $f_{<1}=3\pm1$\% ($3\pm1$\%) in the core and $f_{<1}=12\pm2$\% ($18\pm6$\%) in the 
wings. If we consider the undecomposed total spectrum, we find that the stellar population fraction with 
ages $<1$~Gyr is $f_{<1}=5\pm1$\%. This corresponds to an average SFR of $5.6\pm1$~M$_{\odot}$~yr$^{-1}$ 
over the last 1~Gyr. We note that the actual time resolution of the spectral modeling is significantly 
coarser than the 0.1~Gyr step we use here. Since the fit does not take the [OII]3727$\AA$~line or the 
rest-frame far-UV photometry into account, it is not expected to yield very young solutions.\\

The presence of this additional young stellar population in the wings generates a mild age gradient, 
with the best-fit models to the core and wings having (mass-weighted) average ages of 
$2.8~(2.9)\pm0.1$~Gyr and $2.4~(2.7)\pm0.3$~Gyr for the V10 (M11) templates. These values are summarized 
in Appendix~\ref{spectable}. 
As the V10 templates cover a larger metallicity range than the M11 ones, the solutions they yield tend 
to be smoother in age-space. Nevertheless, in both cases the $>$1~Gyr stellar populations have 
$\log(Z/Z_{\odot})\geq-1$~while the fit yields $\log(Z/Z_{\odot})\geq0$~for the $<$1~Gyr ones, 
similar to the luminosity-weighted value derived from the full stack. 
The wavelength range of the M11 templates also allows us to model the spectrum of the $z>1.5$~pBzKs. 
However, in this case, the only prominent features are the Mg II and Mg I absorption lines at 2800$\AA$ 
and 2852$\AA$, whose behavior is both complex and subject to the well-known age-metallicity degeneracy 
\citep[e.g.,][]{Fan90}. Furthermore, the stack of these higher redshift galaxies has lower S/N and the 
decomposition only yields a usable spectrum for the core (for which we find a mass-weighted age of 
$2\pm1$~Gyr), ruling out a comparison as described above.\\

\section{\label{sfr}Star formation rates}

\subsection{\label{o2}[OII]3727$\AA$~emission}

We subtract the best-fit model to the full spectrum from the observed flux and fit the residuals at 
3727~$\AA$~with a (blended) double Gaussian. We thus estimate the total flux of the [OII]3727$\AA$~line 
at $(3.4\pm0.3)\times10^{-18}$~erg s$^{-1}$ cm$^{-2}$, with $(1.9\pm0.2)\times10^{-18}$~erg s$^{-1}$ 
cm$^{-2}$~in the core and $(1.5\pm0.2)\times10^{-18}$~erg s$^{-1}$ cm$^{-2}$~in the wings (here we only 
consider the core and wings spectra obtained with the Gaussian decomposition of the profile, since both 
parameterizations yield very similar results). This corresponds to equivalent widths (EW) of $5\pm1$~and 
$11\pm2$~for the core and wings, respectively. 
The larger EW of the [OII]3727$\AA$~emission in the wings spectrum (this remains the case even if we 
compare the spectra in inner and outer aperture rather than the core and wings) implies that it is not 
particularly point-like but rather appears to be slightly more extended than the distribution of stellar 
light in the blue optical. This is not very surprising, as we have discarded AGN candidates from this 
subsample. 
We also create a stacked image using cutouts from the UltraVISTA $K_s$-band mosaic, centered at the 
position of the 13 pBzKs in the subsample. The $K_s$-band traces the stellar mass more than the 
rest-frame UV continuum, although at this redshift it can still be biased by young stellar light. 
Using the same stars mentioned in Sect.~\ref{decomp}, we find that the $K_s$-band image and our VIMOS 
spectra have virtually identical spatial resolution. We then compare the $K_s$-band surface brightness 
distribution of the galaxies with the cross-dispersion profile of the [OII]3727$\AA$~line. As shown in 
Fig.~\ref{fig:stelpop}, the latter appears consistent with, although slightly less concentrated than, 
the $K_s$-band profile. Likewise, the cross-dispersion profile of the stellar continuum on both sides 
of the [OII]3727$\AA$~emission is indistinguishable from that of the line. 
We are therefore confident that the ionizing photons giving rise to the [OII]3727$\AA$~emission seen 
in the stacked spectrum indeed originates from young stars (i.e., from star formation), rather than 
from a significant population of hot post-main sequence stars or from nuclear activity. \\

We correct for the loss of flux due to the narrow slit by scaling the median best-fit model to the 
SEDs of pBzKs to the flux of the observed spectrum in the range $3550-4050$~$\AA$. We then multiply 
this normalization factor by the median difference between aperture and total $K_s$-band magnitudes 
in the \citet{Muz13} UltraVISTA catalog to recover the total flux.
Next, we correct this [OII]3727$\AA$~flux for dust extinction using the median value derived from the 
SED fit, $E(B-V)=0.12$, assuming a \citet{Cal00} extinction law and a value of $f=0.83$~for the ratio 
between stellar and nebular extinction \citep[derived for $z\sim1.55$~star-forming galaxies]{Kas13}. 
We convert the [OII]3727$\AA$~flux into a SFR using the relation of \citet{Ken98} and find 
$\text{SFR}_{\text{OII}}=4.5\pm1~(3.7\pm1)$~M$_{\odot}$~yr$^{-1}$ after (before) correcting for 
dust extinction. These values are consistent with the SFH inferred from the spectral modeling. 
Applying the same aperture-to-total correction to the stellar masses, the derived SFR corresponds to a 
specific SFR (sSFR) of $\sim2.7\times10^{-11}$~yr$^{-1}$, $\sim$1.6~dex below the main sequence of star 
formation (MS) at $z=1.4$, assuming the parameterization of \citet[hereafter S14]{Sar14}. This is, 
however, 1~dex higher than the offset between local ETGs and the MS at 
$z\sim0$~\citep[hereafter M16]{Lia16,Man16}. Using the \citet{Cal00} nebular-to-continuum extinction 
ratio ($f=0.44$), on the other hand, would yield 
$\text{SFR}_{\text{OII}}=9\pm2$~M$_{\odot}$~yr$^{-1}$, or $\sim1.2$~dex below the MS.\\

\begin{figure}
\centering
\includegraphics[width=0.5\textwidth]{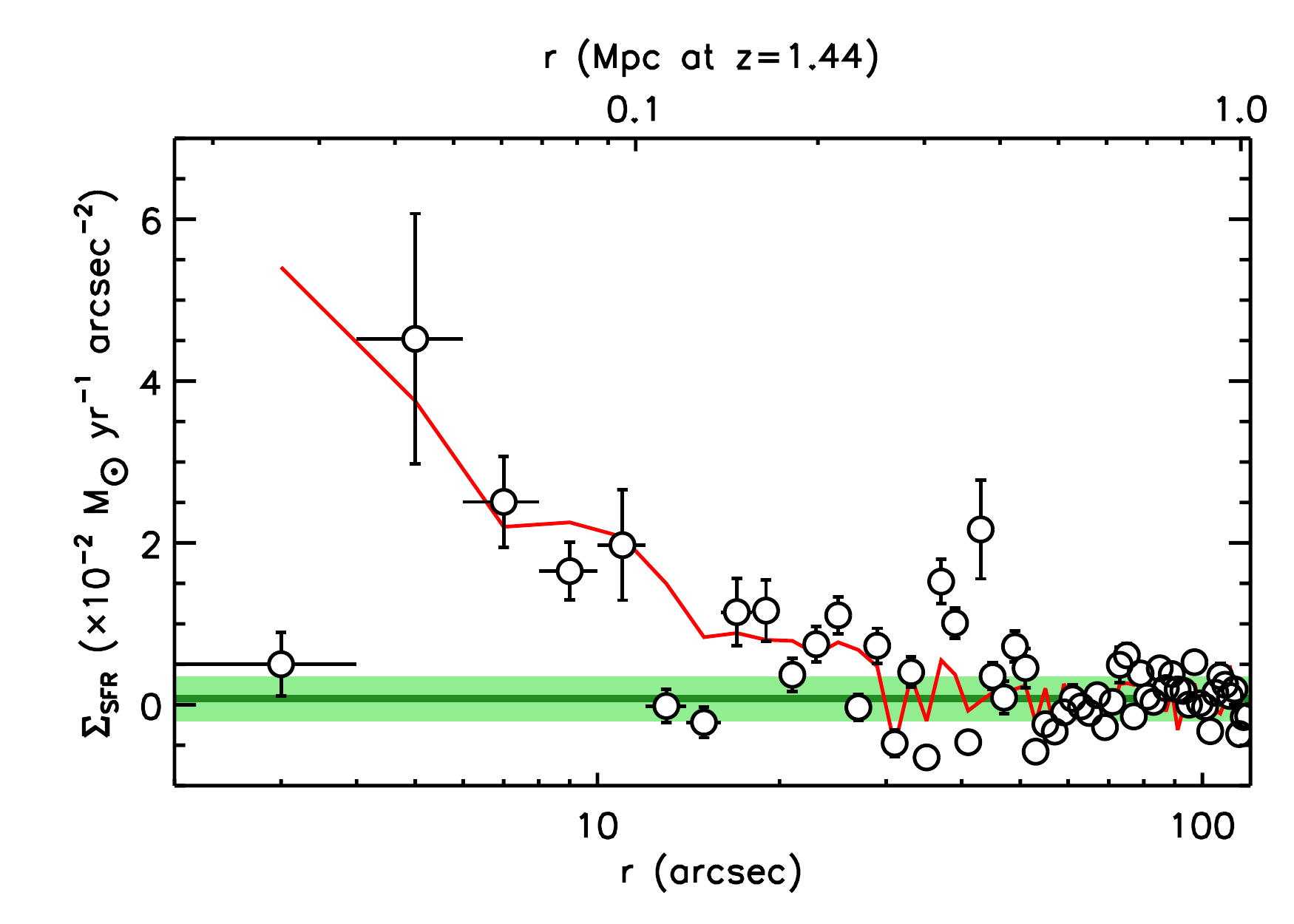}
\caption{Distribution of the SFR density of satellite candidates for the low-z pBzK+UVJp sample, 
shown in bins of 2'' ($\sim$17~kpc at $z=1.44$). The background density, estimated at $r>50''$, is 
shown in green while the red curve shows, for comparison, the total number density of satellite 
candidates, rescaled to match the SFR profile. 
}
\label{fig:sfrsat}
\end{figure}

\subsection{\label{fir}Far-infrared properties}

To further probe the nature of the line emission, we turn to the far-infrared data available on 
the COSMOS field. We extend our spectroscopic sample by including all galaxies with photometric 
redshifts in the range $z=1.2-1.5$, that are selected as passive using both the \emph{BzK} and 
\emph{UVJ} criteria (hereafter pBzK+UVJp) and have $\log M_{\star}\geq 10.8$, to match the stellar 
mass distribution of the spectroscopic sample. We also consider a larger sample consisting of all 
pBzK+UVJp in the COSMOS field, with the same mass cut. We refer to the first one as the ``low-z'' 
sample and the second as the ``total'' one. In both cases, we exclude objects that are individually 
detected at 3$\sigma$ at 24~$\mu$m, using a new catalog of \emph{Spitzer}/MIPS sources from Jin et al. 
(in prep.). We also ignore all possible pairs, i.e., galaxies with concordant redshifts (within 
$\Delta z\leq0.2$) and separated by less than $60$''. This criterion yields 182 pBzK+UVJp galaxies 
in the low-z sample and 977 ones in total sample. Both samples have a median stellar mass of 
$1.2\times10^{11}$~M$_{\odot}$, for a median redshift of $z=1.44$~and $z=1.76$, respectively. 
We then use MIR, FIR, and radio maps, at 24~$\mu$m \citep[\emph{Spitzer}/MIPS; ][]{LeF09}, 
100 and 160~$\mu$m \citep[\emph{Herschel}/PACS, from the  PEP survey;][]{Lut11}, 250,  
350 and 500~$\mu$m \citep[\emph{Herschel}/SPIRE, from the HerMES survey;][]{Oli12}, 
850~$\mu$m \citep[JCMT/SCUBA2;][]{Gea16}, and 1.4~GHz \citep[VLA;][]{Sch10}.\\

\begin{figure}
\centering
\includegraphics[width=0.5\textwidth]{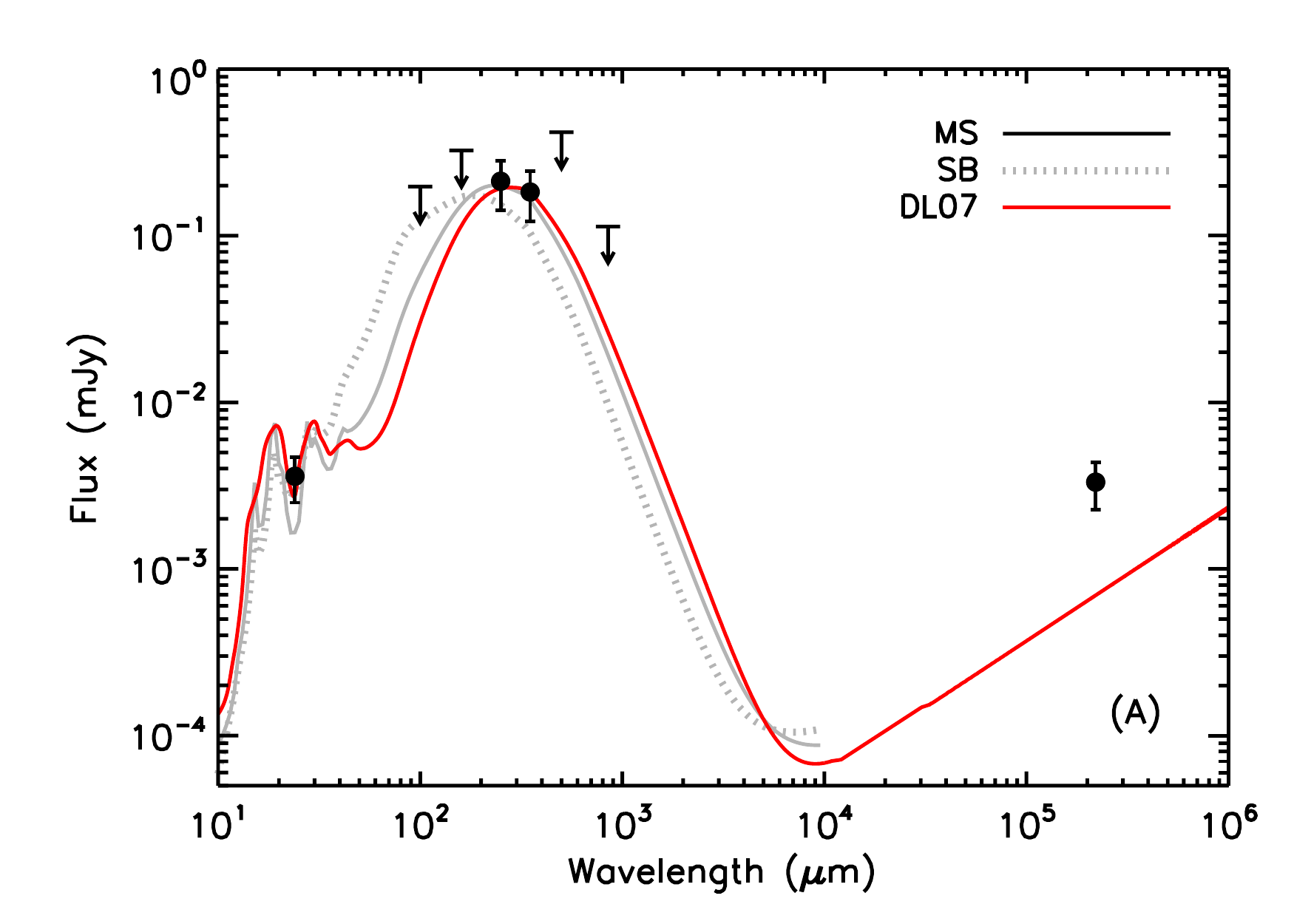}
\includegraphics[width=0.5\textwidth]{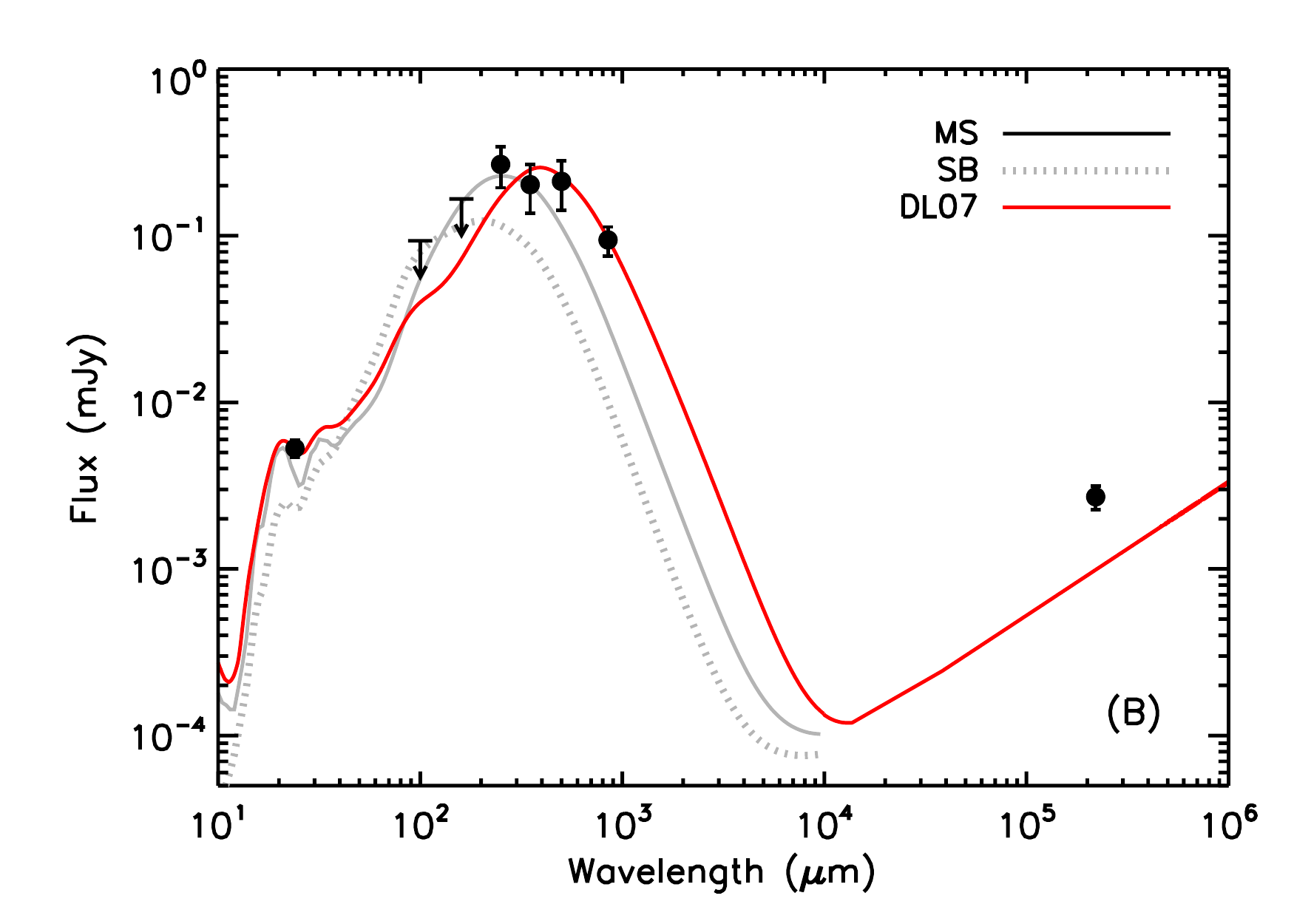}
\caption{
Mid-infrared to radio SED of 24~$\mu$m-undetected pBzK+UVJp galaxies, in the low-z (A) 
and total sample (B), from stacked cutout images (filled circles with error bars, the arrows 
representing 3$\sigma$~upper limits), with best-fit models from \citet{Mag12} (solid and dotted 
gray curves for main sequence and starburst templates, respectively) and \citet{DL07} (solid 
red curve extending to the radio regime). Both the M12 and DL07 models have been broadened 
according to the redshift distributions of each sample.
}
\label{fig:irfit}
\end{figure}

For each band, we create a median 2D image from cutouts around each pBzK+UVJp galaxy 5 times larger 
than the beam FWHM, which we then fit with a centrally positioned point source. This procedure is 
similar to the one described in \citet[hereafter G15]{Gob15} and we refer to that paper for additional 
information. However, we do not include an extended component to the fit, as done in G15, for two reasons: 
first, we are only interested here in the signal coming from the central galaxy; second, unlike for the 
star-forming centrals discussed in G15, the distribution of infrared flux originating from 
the galaxies' satellites is not easily modeled by a simple profile. As shown in 
Fig.~\ref{fig:sfrsat}, the SFR density of satellite candidates falls to the background level close 
to the central, hinting at a strong environmental effect.
Therefore, we apply instead the clustering corrections of \citet[hereafter B15]{Bet15} to the 
measured fluxes, which were derived from simulations using the same COSMOS field. However, since 
B15 did not distinguish between galaxy types, we add an additional correction to the FIR 
bands to account for the increasing fraction of flux from satellites (Fig.~\ref{fig:sfrsat}) 
enclosed by the PACS, SPIRE, and SCUBA2 beams. This effectively raises the clustering corrections 
by up to a factor $\sim$2 at 500~$\mu$m. 
For both samples the stacked cutouts yield strong detections at 24~$\mu$m and 1.4~GHz. We also 
detect the total sample at 3$\sigma$ in the SPIRE and SCUBA2 bands, and the low-z only at 250 and 
350~$\mu$m. The resulting SEDs are shown in Fig.~\ref{fig:irfit}.\\

We then fit the MIR to radio fluxes, including the upper limits, with two sets of dust emission 
models. We use the evolving \citet[hereafter M12]{Mag12} main sequence and starburst templates 
(SB), and the models of \citet[hereafter DL07]{DL07}. For the latter, we set $q_{\text{PAH}}=2.5$\% 
and $U_{\text{max}}=10^6$, but let $U_{\text{min}}$~and $\gamma$~vary. We also extend the models 
to the radio regime, assuming a power law slope with an index of $\alpha=0.8$~and a normalization 
given by the FIR-radio correlation \citep{Mgn15}. The best-fitting models are shown in 
Fig.~\ref{fig:irfit}.
With the MS template, we find 
$\text{SFR}_{\text{MS}}=2.5\pm1.5~(4.3\pm1)$~M$_{\odot}$~yr$^{-1}$~for the low-z (respectively, 
total) sample. We integrate the dust SED (without radio slope) of the DL07 models between 8 
and 1000~$\mu$m to estimate a total infrared luminosity of 
$L_{\text{IR}}=1.3^{+1.2}_{-0.7}~(2.5^{+1.5}_{-0.9})\times10^{10}$~L$_{\odot}$ for the low-z 
(respectively, total) sample, consistent with the absorbed luminosity 
$L_{\text{abs}}\sim3\times10^{10}$~L$_{\odot}$ in the UV, optical, and NIR, which we estimate 
from the best-fit model to the SED and the reddening curve between 90~$\AA$~and 2.5~$\mu$m 
\citep[see][for details]{Cpo16}. We convert $L_{\text{IR}}$ into a SFR using the \citet{Ken98} 
relation and find 
$\text{SFR}_{\text{DL07}}=2.1^{+2.0}_{-1.0}~(4.1^{+2.5}_{-1.5})$~M$_{\odot}$~yr$^{-1}$, or 
1.7 (1.5)~dex below the MS at these redshifts. 
These values are in agreement with the total SFR inferred from the corrected line emission 
as well as with the values derived at $z\sim1-1.5$~by M16 from \emph{Herschel} data, both 
using samples of (differently-selected) quiescent galaxies. 
We also use the FIR SEDs and DL07 modeling thereof to estimate dust masses and temperatures, 
as well as gas masses for both samples. We present and discuss these results elsewhere (Gobat 
et al. 2017, in prep.).
\\

\section{\label{disc}Discussion}

Our analysis suggests the presence of younger stellar populations in the outer regions of the 
pBzK galaxies in our sample compared to their cores. It provides a spectroscopic confirmation of 
the radial color gradients seen in resolved photometry of $z>1$~ETGs \citep{Guo11,Gar12,Cha16}, 
and suggests that they arise at least in part from a radial age gradient. 
While the observed differences in the strength of the 4000~$\AA$ break and the ratio 
of the Ca II H and Ca II K+H$\epsilon$~lines (Fig.~\ref{fig:wings}) between the core (inner) and 
wings (outer) spectra could in principle also arise from a gradient in stellar metallicity, the 
presence of less metal-blended, high-order Balmer lines such as H6 offers a way to partially lift 
this degeneracy \citep[e.g.,][]{vD03,Dem10}. Consequently, the spectral modeling favors a younger 
stellar population in the galaxies' outskirts compared to their cores. 
This stellar population difference can be interpreted in different ways: on the one hand, this could 
be a signature of the recent accretion of low-mass satellites with younger stellar populations by 
this galaxy population. In this case, star formation would likely occur in extended or disk-like 
features \citep[e.g.,][Mancini et al., in prep.]{App14}, consistent with our observations. 
Since this stellar population ratio is derived using a median spectrum, this would then imply 
that most of the galaxies contributing to the stack experienced such a rejuvenation episode during 
the last 1~Gyr.\\

On the other hand, it could be seen as tentative evidence for an inside-out 
quenching of star formation in these galaxies. This would be consistent with observations of depressed 
star formation in the inner regions of $z\gtrsim2$~massive MS galaxies \citep{Wuy13,Gen14,Tac15}, 
which are potential progenitors to the $z<1.5$~ETGs.
However, the young stellar population contributes only a relatively small fraction of the total 
stellar mass in either the core or wings. From the results of the spectral modeling, we can thus 
constrain the age gradient to be $<$0.5~Gyr over $\sim$6~kpc ($\sim$4~pixels) between the wings and 
core, respectively. This is a somewhat smaller value than would be expected from pure inside-out 
quenching 
\citep[$\sim$2~Gyr over the same range for a $\sim$10$^{11}$~M$_{\odot}$~galaxy; e.g.,][]{Mrt09,Tac15}. 
It implies a faster quenching timescale \citep[see, e.g.,][for the case of $z\sim1.6$~ETGs]{Ono15} 
and thus an additional quenching mechanism. Indeed, the apparent SFR deficit in close satellites of 
the pBzK+UVJp galaxies, as shown in Fig.~\ref{fig:sfrsat}, suggests that their immediate 
environment is efficient at suppressing star formation, providing further evidence that even at 
$z>1$~the host halo could contribute an important quenching channel for high mass galaxies. 
Finally, we note that this small age gradient is of the same order as the ones observed in local 
massive ETGs \citep[e.g.,][]{Spo10,Kun10,Kol11,Bar16}, consistent with a mostly passive evolution 
to the present day.\\

We also detect [OII]3727$\AA$~emission in the stacked spectrum. Although we have taken steps to 
remove AGN candidates from the stack, a low-ionization nuclear emission region (LINER) could still be 
a possible source for it, considering the nature of the sample \citep[e.g.,][]{Yan06}. 
On the other hand, the [OII]3727$\AA$~emission is distinctively more spatially extended than a point 
source, as shown in Fig.~\ref{fig:stelpop}. We can thus interpret it as arising from low-level star 
formation consistent with the reconstructed SFH of the pBzK galaxies. 
Since we use a median stack, it is quite unlikely that this emission comes from separate but 
unresolved satellites rather than the galaxies themselves. If the age difference between the core and 
wings spectra is indeed a signature of inside-out quenching, one would expect the sSFR estimated 
from the line emission to be lower in the core than in the wings. However we cannot presently test this 
hypothesis: while the spatial profile of the line does appear to be slightly broader than the average 
$K_s$-band profile of the galaxies, suggesting a deficit of [OII]3727$\AA$~emission per unit stellar 
mass in the core, the relatively low spatial resolution of both the broad-band image and our VIMOS 
spectra precludes an unambiguous comparison. Since our spectroscopic pBzK sample does not overlap 
with the CANDELS survey \citep{Koe11}, we do not have access to precise NIR morphologies for these 
galaxies nor, as a consequence, to truly resolved stellar mass information.\\ 

The SFR estimated from the [OII]3727$\AA$~emission is consistent with both the stacked MIR and FIR 
emission from a larger sample of similarly-selected quiescent galaxies. This disagrees with the 
results of M16, who find that the SFRs derived from MIR fluxes are generally lower than the ones 
estimated from the FIR. Conversely, \citet{Fum14} conclude that the heating of cirrus dust by old 
stellar populations, rather than SFR, dominates the MIR emission of $z\sim1.5$~quiescent galaxies by 
an order of magnitude.
Similarly, \citet{Hay14} find that, in simulated quenched galaxies, $L_{\text{IR}}$ tends to overestimate 
the true SFR by an order of magnitude for as long as 1~Gyr, due to residual heating of the dust by the 
last batch of stars. However, since the different indicators used here probe different timescales (a few 
tens of Myr in the case of line emission) the good agreement between SFR estimates derived from 
[OII]3727$\AA$ and \emph{Herschel} suggests that these alternative explanations are unlikely.\\

Finally, we note that the best-fit model underestimates the radio fluxes by a factor $\sim3-5$ in 
both cases, which suggests the presence of low-level AGN activity in this sample \citep[e.g.,][]{Sar10}. 
The excess radio luminosity is $\sim3\times10^{22}$~W/Hz, which falls well within the range dominated 
by low-excitation (radio-mode) AGNs \citep{BH12}, even accounting for a non-zero spectral index (and 
therefore K-correction). This type of AGN is commonly found locally in massive red-sequence galaxies 
\citep[e.g.,][]{Smo09} and associated with ``maintenance mode'' feedback, where the AGN outflow heats 
the surrounding extra-galactic medium, therefore preventing star formation. 
This excess is also present if we remove from the stack the cutouts that fall within the lower and upper 
16 percentiles of the flux within a central aperture, thus removing potential biases from both negative 
noise peaks and detected radio sources. This would suggest that radio-mode feedback is pervasive up to 
at least $z\sim1.8$, possibly playing a role in keeping the star formation rate low in these galaxies.\\

\section{\label{sum}Summary}

We present VLT/VIMOS spectroscopic observations of a sample of 35 massive \emph{BzK}-selected quiescent 
galaxies in the COSMOS field. We derive reliable redshifts for 31 sources, which we use together with 
available multiband photometry to estimate stellar masses and dust extinction values. Our sample has a 
median redshift of $z=1.51$~and a median stellar mass of $1.1\times10^{11}$~M$_{\odot}$~(Salpeter IMF). 
Our findings are summarized as follows:\\

$\bullet$~we create a median, spatially resolved 2D spectrum of the securely confirmed pBzK galaxies. 
We find that the outer part of this composite spectrum has features consistent with $f_{<1}\gtrsim12$\% 
of its stellar population being $<$1~Gyr old, against $f_{<1}\lesssim4$\% in the inner part. This 
corresponds to an age difference of $\leq$0.4~Gyr over 6~kpc. We interpret this as possible evidence 
for either wide-spread rejuvenation episodes or an inside-out quenching process in these galaxies.\\

$\bullet$~we detect extended [OII]3727$\AA$~emission consistent with residual star formation at a rate 
of $\text{SFR}_{\text{OII}}\sim3-5$~M$_{\odot}$~yr$^{-1}$, after correcting for extinction, aperture, 
and slit losses.\\ 

$\bullet$~we perform stacked MIR, FIR, and radio photometry on a larger sample, which includes 
all galaxies in the COSMOS field in the same mass range that are selected as passive by both the 
\emph{BzK}~and \emph{UVJ} criteria and are undetected in the MIR. Our analysis yields FIR detections 
that allow us to derive a $\text{SFR}_{\text{IR}}$~of $2-4$~M$_{\odot}$~yr$^{-1}$, consistent with 
the extinction-corrected value derived from the [OII]3727$\AA$~emission. On the other hand, we find 
radio fluxes a factor $\sim3-5$ higher than predicted by the SFR, implying the widespread 
presence in the photometric sample of low-luminosity AGNs associated with maintenance-mode feedback.\\

$\bullet$~given the median stellar masses of the spectroscopic and photometric samples, we 
find that quiescent galaxies at $z=1.4-1.8$~have SFRs $1.5-1.7$~dex below the MS at these 
redshifts. This is $\sim$1~dex higher than the offset reported for local ETGs 
($\sim-$2.5~dex).\\

\begin{acknowledgements}
AC acknowledges the grant MIUR PRIN 2010-2011 "The dark Universe and the cosmic evolution of baryons: 
from current surveys to Euclid". MC acknowledges support from a Royal Society University Research 
Fellowship, MTS from a Royal Society Leverhulme Trust Senior Research Fellowship. CF acknowledges 
funding from the European Union's Horizon 2020 research and innovation programme under the Marie 
Sk\l{}odowska-Curie grant agreement No 664931. GEM acknowledges support from the ERC Consolidator 
Grant funding scheme (project ConTExt, grant number No. 648179) and a research grant (13160) from 
Villum Fonden.
\end{acknowledgements}

\begin{appendix}

\section{\label{ztable}Spectroscopic redshifts}

\begin{table}
\caption{Spectroscopic redshifts of the pBzK galaxies in our sample. 
Column (1): galaxy ID in the \citet{Muz13} catalog; column (2) and (3): 
right ascension and declination (J2000 epoch), from the $K$-band selected 
catalog; column (4): spectroscopic redshift; column (5): quality flag.}
\centering
\setlength{\tabcolsep}{10pt}
\begin{tabular}{c c c c c}
\hline\hline
ID & RA & DEC & $z_{\text{spec}}$ & Quality\\
& (deg) & (deg) & & \\
\hline
206430 & 150.00462 & 2.47327 & 1.4595 & A \\
207538 & 150.01192 & 2.48666 & 1.5499 & A \\
208020 & 150.04401 & 2.49370 & 1.9677 & B$^-$ \\
211123 & 150.04243 & 2.52651 & 1.0296 & C \\
214127 & 150.09035 & 2.56794 & 1.9677 & B \\
218368 & 150.09189 & 2.61840 & 1.6658 & B$^-$ \\
219830 & 150.09418 & 2.63460 & 1.3965 & A \\
220268 & 150.01241 & 2.64065 & 2.0892 & B \\
221046 & 150.04456 & 2.65100 & 1.4408 & A \\
221493 & 150.00569 & 2.65488 & 1.2471 & A \\
222897 & 150.01764 & 2.67258 & 1.5839 & A \\
223635 & 150.00270 & 2.67492 & 1.2468 & A \\
223796 & 149.98267 & 2.68305 & 1.4372 & A \\
225039 & 150.08479 & 2.69475 & 1.3005 & A \\
225317 & 150.10361 & 2.69767 & 1.4072 & A \\
226037 & 150.08859 & 2.70642 & 1.3770 & C \\
220655 & 150.00571 & 2.64554 & 0.8938 & B \\
221824 & 150.01912 & 2.65889 & 1.3961 & A \\
226380 & 150.09518 & 2.71000 & 1.4122 & B \\
248681 & 149.89543 & 2.62726 & 1.5596 & A \\
248881 & 149.81070 & 2.63236 & 1.5199 & A \\
250123 & 149.86220 & 2.64795 & 1.5681 & B \\
250513 & 149.82225 & 2.65310 & 2.0799 & B \\
252409 & 149.86949 & 2.67430 & 1.4050 & A \\
254526 & 149.80560 & 2.70258 & 1.5126 & A \\
254313 & 149.87495 & 2.69825 & 1.3149 & A \\
236990 & 149.88060 & 2.48648 & 1.5519 & A \\
238164 & 149.89635 & 2.49910 & 1.5642 & A \\
240680 & 149.83612 & 2.52852 & 1.6587 & C \\
243072 & 149.89896 & 2.55742 & 1.5888 & B \\
243201 & 149.89194 & 2.56043 & 1.5939 & A \\
243731 & 149.82497 & 2.56692 & 1.5145 & B \\
237270 & 149.85873 & 2.48966 & 1.8368 & B \\
241810 & 149.85217 & 2.54303 & 1.1757 & A \\
\hline
\end{tabular}
\end{table}

\section{\label{spectable}Stellar population properties}

\begin{table}
\caption{Results of the spectral modeling for the different types of 
spectral decomposition (column 1) and models (column 2). Column (3) and 
(4): fraction of stars with ages $<$1~Gyr in the core and wings, respectively; 
column (5) and (6): mass-weighted average age of the core and wings, respectively}
\centering
\bgroup
\setlength{\tabcolsep}{4pt}
\def\arraystretch{1.5}
\begin{tabular}{c c c c c c}
\hline\hline
& & $f_{<1,\text{core}}$ & $f_{<1,\text{wings}}$ & 
age$_{\text{core}}$ & age$_{\text{wings}}$ \\
& & (\%) & (\%) & (Gyr) & (Gyr)\\
\hline
Gaussian & V10 & $3\pm1$ & $13\pm1$ & $2.8\pm0.1$ & $2.4\pm0.3$\\
& M11 & $4\pm1$ & $21\pm7$ & $2.9\pm0.1$ & $2.7\pm0.3$\\
de Vaucouleurs & V10 & $3\pm1$ & $12\pm2$ & $2.8\pm0.1$ & $2.4\pm0.3$\\
& M11 & $3\pm1$ & $18\pm6$ & $2.9\pm0.7$ & $2.7\pm0.9$\\
\hline
\end{tabular}
\egroup
\end{table}

\end{appendix}


\begin{thebibliography}{1}

\bibitem[Appleton et al.(2014)]{App14}Appleton, P.N. et al., 2014, \apj, 797, 117
\bibitem[Baldry et al.(2006)]{Bal06}Baldry, I.K. et al., 2006, \mnras, 373, 469
\bibitem[Barbosa et al.(2016)]{Bar16}Barbosa, C.E. et al., 2016, \aap, 589, 139
\bibitem[Belli et al.(2014)]{Bel14}Belli, S. et al., 2014, \apjl, 788, 29
\bibitem[Best \& Heckman(2012)]{BH12}Best, P.N., Heckman, T.M., 2012, \mnras, 421, 1569
\bibitem[B\'{e}thermin et al.(2015)]{Bet15}B\'{e}thermin, M. et al., 2015, \aap, 573, 113
\bibitem[Birnboim \& Dekel(2003)]{Bir03}Birnboim, Y, Dekel, A., 2003, \mnras, 345, 349
\bibitem[Bruzual \& Charlot(2003)]{BC03}Bruzual, G. \& Charlot, S., 2003, MNRAS, 344, 1000
\bibitem[Calzetti et al.(2000)]{Cal00}Calzetti, D. et al., 2000, \apj, 533, 682
\bibitem[Capozzi et al.(2016)]{Cpo16}Capozzi, D. et al., 2016, \mnras, 456, 790
\bibitem[Cappellari et al.(2013a)]{Cap13a}Cappellari, M. et al., 2013a, \mnras, 432, 1709
\bibitem[Cappellari et al.(2013b)]{Cap13b}Cappellari, M. et al., 2013b, \mnras, 432, 1862
\bibitem[Carollo et al.(2013)]{Car13}Carollo, C.M. et al., 2013, \apj, 773, 112
\bibitem[Ceverino \& Klypin(2009)]{Cev09}Ceverino, D., Klypin, A., 2009, \apj, 695, 292
\bibitem[Chan et al.(2016)]{Cha16}Chan, J.C.C. et al., 2016, \mnras, 458, 3181
\bibitem[Chiang et al.(2014)]{Chi14}Chiang, Y.-K. et al., 2014, \apjl, 782, 3
\bibitem[Choi et al.(2014)]{Cho14}Choi, J. et al., 2014, \apj, 792, 95
\bibitem[Ciesla et al.(2014)]{Cie14}Ciesla, L. et al., 2014, \aap, 565, 128
\bibitem[Cimatti et al.(2002)]{Cim02}Cimatti, A. et al., 2002, \aap, 381L, 68
\bibitem[Cimatti et al.(2004)]{Cim04}Cimatti, A. et al., 2004, Nature, 430, 184
\bibitem[Citro et al.(2016)]{Cit16}Citro, A. et al., 2016, \aap, 592, 19
\bibitem[Croton et al.(2006)]{Crt06}Croton, D.J. et al., 2006, \mnras, 365, 11
\bibitem[Daddi et al.(2004)]{Dad04}Daddi, E. et al., 2004, \apj, 617, 746
\bibitem[Daddi et al.(2005)]{Dad05}Daddi, E. et al., 2005, \apj, 626, 680
\bibitem[Daddi et al.(2010)]{Dad10}Daddi, E. et al., 2010, \apjl, 714, 118
\bibitem[Davidge \& Clark(1994)]{Dav94}Davidge, T.J., Clark, C.C., 1994, \aj, 107, 946
\bibitem[Dekel \& Birnboim(2006)]{Dek06}Dekel, A. \& Birnboim, Y., 2006, \mnras, 368, 2
\bibitem[Demarco et al.(2010)]{Dem10}Demarco, R. et al., 2010, \apj, 725, 1252
\bibitem[Di Matteo et al.(2005)]{DiM05}Di Matteo, T. et al., 2005, \nat, 433, 604
\bibitem[Donley et al.(2012)]{Don12}Donley, J.L. et al., 2012, \apj, 748, 142
\bibitem[Draine \& Li(2007)]{DL07}Draine, B.T., Li, A., 2007, \apj, 657, 810
\bibitem[Fanelli et al.(1990)]{Fan90}Fanelli, M.N. et al., 1990, \apj, 364, 272
\bibitem[Feldmann \& Mayer(2015)]{Fel15}Feldmann, R., Mayer, L., 2015, \mnras, 446, 1939
\bibitem[Fumagalli et al.(2014)]{Fum14}Fumagalli, M. et al., 2014, \apj, 796, 35
\bibitem[Gabor et al.(2010)]{Gab10}Gabor, J.M. et al., 2010, \mnras, 407, 749
\bibitem[Gallazzi et al.(2006)]{Gal06}Gallazzi, A. et al., 2006, \mnras, 370, 1106
\bibitem[Gallazzi et al.(2014)]{Gal14}Gallazzi, A. et al., 2014, \apj, 788, 72
\bibitem[Gargiulo et al.(2012)]{Gar12}Gargiulo, A. et al., 2012, \mnras, 425, 2698
\bibitem[Geach et al.(2016)]{Gea16}Geach, J.E. et al., 2016, arXiv:1607.03904
\bibitem[Genzel et al.(2014)]{Gen14}Genzel, R. et al., 2014, \apj, 785, 75
\bibitem[Gobat et al.(2011)]{Gob11}Gobat, R. et al., 2011, \aap, 526, 133
\bibitem[Gobat et al.(2012)]{Gob12}Gobat, R. et al., 2012, \apjl, 759, 44
\bibitem[Gobat et al.(2013)]{Gob13}Gobat, R., et al. 2013, \apj, 776, 9
\bibitem[Gobat et al.(2015)]{Gob15}Gobat, R. et al., 2015, \aap, 581, 56
\bibitem[Granato et al.(2004)]{Gra04}Granato, G.L. et al., 2004, \apj, 600, 580
\bibitem[Guo et al.(2011)]{Guo11}Guo, Y. et al., 2011, \apj, 735, 18
\bibitem[Hayward et al.(2014)]{Hay14}Hayward, C.C. et al., 2014, \mnras, 445, 1598
\bibitem[Hill et al.(2016)]{Hil16}Hill, A.R. et al., 2016, \apj, 819, 74
\bibitem[Hopkins et al.(2006)]{Hop06}Hopkins, P.F. et al., \apjs, 163, 50
\bibitem[Kashino et al.(2013)]{Kas13}Kashino, D., 2013, \apjl, 777, 8
\bibitem[Kauffmann et al.(2014)]{Kau04}Kauffmann, G. et al., 2004, \mnras, 353, 713
\bibitem[Kennicutt(1998)]{Ken98}Kennicutt, R.C., 1998, \araa, 36
\bibitem[Koekemoer et al.(2007)]{Koe07}Koekemoer, A.M. et al., 2007, \apjs, 172, 196
\bibitem[Koekemoer et al.(2011)]{Koe11}Koekemoer, A.M. et al., 2011, \apjs, 197, 36
\bibitem[Koleva et al.(2011)]{Kol11}Koleva, M. et al., 2011, \mnras, 417, 1643
\bibitem[Krogager et al.(2014)]{Kro14}Krogager, J.-K. et al., 2014, \apj, 797, 17
\bibitem[Kuntschner et al.(2010)]{Kun10}Kuntschner, H. et al., 2010, \mnras, 408, 97
\bibitem[Kurk et al.(2013)]{Kur13}Kurk, J. et al., 2013, \aap, 549, 63
\bibitem[Le Floc'h et al.(2009)]{LeF09}Le Floc'h, E. et al., 2009, \apj, 703, 222
\bibitem[Lianou et al.(2016)]{Lia16}Lianou, S. et al., 2016, \mnras, 461, 2856
\bibitem[Lonoce et al.(2014)]{Lon14}Lonoce, I et al., 2014, \mnras, 444, 2048
\bibitem[Lutz et al.(2011)]{Lut11}Lutz, D. et al., 2011, \aap, 532, 90
\bibitem[Magdis et al.(2012)]{Mag12}Magdis, G. et al., 2012, \apj, 760, 6
\bibitem[Magnelli et al.(2015)]{Mgn15}Magnelli, B. et al., 2015, \aap, 573, 45
\bibitem[Man et al.(2016)]{Man16}Man, A.W.S. et al., 2016, \apj, 820, 11
\bibitem[Mancini et al.(2010)]{Mni10}Mancini, C. et al., 2010, \mnras, 401, 933
\bibitem[Maraston \& Str\"{o}mb\"{a}ck(2011)]{M11}Maraston, C., Str\"{o}mb\"{a}ck, G., 
2011, \mnras, 418, 2785
\bibitem[Markwardt(2009)]{Mar09}Markwardt, C.B., 2009, ASPC, 411, 251
\bibitem[Martig et al.(2009)]{Mrt09}Martig, M. et al., 2009, \apj, 707, 250
\bibitem[Martig et al.(2013)]{Mrt13}Martig, M. et al., 2013, \mnras, 432, 1914
\bibitem[McCarthy et al.(2004)]{MCy04}McCarthy, P.J. et al., 2004, \apj, 614, 9
\bibitem[McCracken et al.(2010)]{McC10}McCracken, H.J. et al., 2010, \apj, 708, 202
\bibitem[McCracken et al.(2012)]{McC12}McCracken, H.J. et al., 2012, \aap, 544, 156
\bibitem[Mellier \& Mathez(1987)]{MM87}Mellier, Y., Mathez, G., 1987, \aap, 175, 1
\bibitem[Muzzin et al.(2013)]{Muz13}Muzzin, A. et al., 2013, \apjs, 206, 8
\bibitem[Newman et al.(2014)]{New14}Newman, A.B. et al., 2014, \apj, 788, 51
\bibitem[Oliver et al.(2012)]{Oli12}Oliver, S.J. et al. 2012, \mnras, 424, 1614
\bibitem[Onodera et al.(2012)]{Ono12}Onodera, M. et al., 2012, \apj, 755, 26
\bibitem[Onodera et al.(2015)]{Ono15}Onodera, M. et al., 2015, \apj, 808, 161
\bibitem[Peng et al.(2010)]{Pen10}Peng, Y.-J. et al., 2010, \apj, 721, 193
\bibitem[Rettura et al.(2010)]{Ret10}Rettura, A. et al., 2010, \apj, 709, 512
\bibitem[Rettura et al.(2011)]{Ret11}Rettura, A. et al., 2010, \apj, 732, 94
\bibitem[Saintonge et al.(2012)]{Sai12}Saintonge, A. et al., 2012, \apj, 758, 73
\bibitem[Salpeter(1955)]{Sal55}Salpeter, E.E., 1955, \apj, 121, 161
\bibitem[S\'{a}nchez-Bl\'{a}zquez et al.(2006)]{SB06}S\'{a}nchez-Bl\'{a}zquez, P. et 
al., 2006, \mnras, 371, 703
\bibitem[Sargent et al.(2010)]{Sar10}Sargent, M. et al., 2010, \apjs, 186, 341
\bibitem[Sargent et al.(2014)]{Sar14}Sargent, M. et al., 2014, \apj, 793, 19
\bibitem[Sargent et al.(2015)]{Sar15}Sargent, M. et al., 2015, \apjl, 806, 20
\bibitem[Schinnerer et al.(2010)]{Sch10}Schinnerer, E. et al., 2010, \apjs, 188, 384
\bibitem[Scodeggio et al.(2005)]{Sco05}Scodeggio, M. et al., 2005, \pasp, 117, 1284
\bibitem[Scoville et al.(2007)]{Sco07}Scoville, N. et al., 2007, \apjs, 172, 1
\bibitem[Serven et al.(2011)]{Ser11}Serven, J. et al., 2011, \aj, 141, 184
\bibitem[Shetti \& Cappellari(2015)]{She15}Shetty, S. Cappellari, M., 2015, \mnras, 
454, 1332
\bibitem[Smol\u{c}i\'{c}(2009)]{Smo09}Smol\u{c}i\'{c}, V., 2009, \apjl, 699, 43
\bibitem[Spitler et al.(2012)]{Spi12}Spitler, L.R. et al., 2012, \apjl, 748, 21
\bibitem[Spolaor et al.(2010)]{Spo10}Spolaor, M. et al., 2010, \mnras, 408, 272
\bibitem[Strazzullo et al.(2015)]{Str15}Strazzullo et al., 2015, \aap, 576L, 6
\bibitem[Tacchella et al.(2015)]{Tac15}Tacchella, S. et al., 2015, Science, 348, 314
\bibitem[Tanaka et al.(2013)]{Tan13}Tanaka, M. et al., 2013, \apj, 772, 113
\bibitem[Thomas et al.(2005)]{Tho05}Thomas, D. et al., 2005, \apj, 621, 673
\bibitem[Thomas et al.(2010)]{Tho10}Thomas, D. et al., 2010, \mnras, 404, 1775
\bibitem[Utomo et al.(2015)]{Uto15}Utomo, D. et al., 2015, \apj, 803, 16
\bibitem[van Dokkum \& Stanford(2003)]{vD03}van Dokkum, P.G., Stanford, S.A., 2003, \apj, 
585, 78
\bibitem[van Dokkum et al.(2008)]{vDk08}van Dokkum, P.G. et al., 2008, \apjl, 677, 5
\bibitem[Vazdekis et al.(2010)]{Vaz10}Vazdekis, A. et al., 2010, \mnras, 404, 1639
\bibitem[Wagner et al.(2015)]{Wag15}Wagner, C.R. et al., 2015, \apj, 800, 107
\bibitem[Whitaker et al.(2013)]{Whi13}Whitaker, K.E. et al., 2013, \apjl, 770, 39
\bibitem[Williams et al.(2009)]{Wil09}Williams, R.J. et al., 2009, \apj, 691, 1879
\bibitem[Williams et al.(2016)]{Wll16}Williams, C.C. et al., 2016, arXiv:1607.06089
\bibitem[Woo et al.(2015)]{Woo15}Woo, J. et al., 2015, \mnras, 448, 237
\bibitem[Wuyts et al.(2013)]{Wuy13}Wuyts, S. et al., 2013, \apj, 779, 135
\bibitem[Yan et al.(2006)]{Yan06}Yan, R. et al., 2006, \apj, 648. 281

\end{thebibliography}
\end{document}